\documentclass[preprint,5p,times]{elsarticle}
\usepackage{amsmath}
\usepackage{booktabs}
\usepackage{siunitx}
\usepackage{algorithmic}
\usepackage[ruled]{algorithm2e}
\usepackage{pgfplots}
\usepackage{url}
\usepackage[font=normalsize]{caption}
\usepackage{subcaption}
\usepackage{multirow}
\usepackage{amssymb}

\def\eg{{e.g.}}

\usepackage{mathtools}

\DeclarePairedDelimiter\floor{\lfloor}{\rfloor}

\begin{document}
	\begin{sloppypar}
		\begin{frontmatter}
			
			\title{RoadFed: A Multimodal \emph{Fed}erated Learning System for Improving \emph{Road} Safety}
			
			\author[1,2]{Yachao~Yuan} 
			\affiliation[1]{organization={School of Future Science and Engineering, Soochow University},
				city={Suzhou},
				state={Jiangsu},
				country={China}}
			
			\affiliation[2]{organization={Key Laboratory of Computer Network and Information Integration (Southeast University), Ministry of Education},
				city={Nanjing},
				state={Jiangsu},
				country={China}}
			
			\affiliation[3]{organization={School of Cyber Science and Engineering, Southeast  University},
				city={Nanjing},
				state={Jiangsu},
				country={China}}
			
			\affiliation[4]{organization={College of Computing and Intelligent Systems, University of Khorfakkan},
				city={Sharjah},
				country={UAE}}
			
			\author[1]{Zhen~Yu} 
			\author[3]{Yali Yuan*} 
			\ead{Co-corresponding authors: (yaliyuan@seu.edu.cn)}
			\author[1]{Xingyu~Chen}
			\author[1,2]{Yingwen~Wu*} 
			\ead{Co-corresponding authors: (ywwu@suda.edu.cn)}
			
			\author[4]{Thar Baker}

			\begin{abstract}
				Internet of Things (IoTs) have been widely applied in Collaborative Intelligent Transportation Systems (C-ITS) for the prevention of road accidents. As one of the primary causes of road accidents in C-ITS, the efficient detection and early alarm of road hazards are of paramount importance. Given the importance, extensive research has explored this topic and obtained favorable results. However, most existing solutions only explore single-modality data, struggle with high computation and communication overhead, or suffer from the curse of high dimensionality in their privacy-preserving methodologies. To overcome these obstacles, in this paper, we introduce RoadFed, an innovative and private multimodal \emph{Fed}erated learning-based system tailored for intelligent \emph{Road} hazard detection and alarm. This framework encompasses an innovative Multimodal Road Hazard Detector, a communication-efficient federated learning approach, and a customized low-error-rate local differential privacy method crafted for high dimensional multimodal data. Experimental results reveal that the proposed RoadFed surpasses most existing systems in the self-gathered real-world and CrisisMMD public datasets. In particular, RoadFed achieves an accuracy of 96.42\% with a mere 0.0351 seconds of latency, and its communication cost is much lower than existing systems in this field. It facilitates collaborative training with non-i.i.d. high dimensional multimodal real-world data across various data modalities on multiple edges while ensuring privacy preservation for road users. 
			\end{abstract}
			
			
			
			
		\end{frontmatter}
		
		\section{Introduction} \label{sec:Introduction}
		Internet of Things (IoTs) has been widely applied in diverse sectors such as smart cities \cite{tan2024privacy}, and Intelligent Transportation Systems (ITSs) \cite{bakirci2025advanced}, bringing huge revolutions in people's lifestyles. In this context, various IoT devices collect massive amounts of data from the real-world environment, providing users with high-quality services through modern digital technologies. Traffic accident prevention in ITS has attracted significant attention in industry and academia. The massive traffic data collected by IoT devices in ITSs is broadly used for traffic problems, such as traffic accident prevention. 
		Since 2010, the annual number of fatalities resulting from traffic accidents has seen a slight decline, reaching 1.19 million and imposing costs on governments equivalent to approximately 1\%-3\% of Gross Domestic Product (GDP), as noted in~\cite{world2023global}. One of the main contributing factors to these incidents is road hazards, which include issues such as damaged roads, fallen trees, and crashed vehicles. However, because of huge road networks, messy real-world backgrounds, and high intra-class differences, it is very challenging for road users to receive useful road hazard information. 
		
		\textcolor{black}{Recent intelligent road hazard recognition frameworks like \cite{Edge-cloudRD,EEFNet} employ edge-based frameworks for fast road damage inspection by placing the detection models at edges.} However, most of them only use single-modality data, while extensive data in other modalities from IoT devices, such as text, remains unexplored. 
		\textcolor{black}{Conventionally, Single-modality data consists of information from one source, such as just text or image. In contrast, multiple-modality data combines information from two or more different sources, like using both an image and its descriptive text to provide a richer understanding.}
		\textcolor{black}{Furthermore, most existing approaches like \cite{Edge-cloudRD,EEFNet} identify road hazards on an edge by a machine learning model trained with large annotated datasets.} However, the gathering of large annotated datasets is laborious, and the model struggles to dynamically update its knowledge based on evolving data patterns.
		
		Federated Learning (FL)~\cite{mcmahan2017communication,zhang2025adaptive} allows different platforms to acquire a global model while maintaining training data locally on road users' devices, providing privacy and security to some extent. \textcolor{black}{Many studies~\cite{EEFNet,zhao2024community,saha2024federated} proposed various FL strategies to improve FedAvg's~\cite{mcmahan2017communication} performance in the application of road damage detection.} 
		Despite ongoing advancements, they still face several persistent challenges.
		One issue is the heterogeneity of data produced by IoT devices across various systems (\eg, non-i.i.d. data, \textcolor{black}{where the data points are not drawn from the same underlying distribution and are not statistically independent of each other}), which complicates the process of deriving meaningful insights. Additionally, the significant communication overhead remains a critical limitation for the practical deployment of federated learning in real-world settings.
		
		Besides, the massive data produced by IoT devices incorporates large amounts of sensitive information, such as location privacy and facial privacy, which can be exposed to the adversary due to frequent bidirectional communications among users, edges, and the cloud. FL mitigates the privacy issue to some extent; however,~\cite{geiping2020inverting} demonstrates that people can still recover private data directly from the shared gradient parameters in FL. Differential Privacy (DP)~\cite{dwork2014algorithmic} is a promising strategy to protect sensitive information while maintaining model performance. \textcolor{black}{Previous research, \cite{EEFNet,xiong2021privacy,saha2024federated}, preserves data privacy at road users' devices, however, most of them are not for ITSs or did not take privacy of multimodal data into consideration. Even though some existing work, like~\cite{EEFNet}, preserves data privacy by using DP, the expected error of their methods is excessively high when handling high dimensional real-world data from IoT devices.} 
		
		To tackle these issues, \emph{RoadFed}: a multimodal \emph{Fed}erated learning system is developed in this paper for improving \emph{Road} safety.
		It capitalizes on the recent achievements in federated learning, edge-cloud computing, and Local Differential Privacy (LDP) to provide distributed and privacy-preserving road condition monitoring and danger alarm. 
		RoadFed notably minimizes latency by identifying road hazards at the edge servers. By integrating visual and textual data for model training, RoadFed achieves superior detection accuracy compared to previous methods. Additionally, it enables collaborative and efficient learning across multiple edges while ensuring that most data remain securely on users’ devices, enhancing privacy. Furthermore, the proposed Multimodal Local Differential Privacy algorithm offers an extra layer of data protection.
		The \textbf{key contributions} of this paper are outlined below:
		\begin{itemize}
			\item 
			\textcolor{black}{A novel edge-cloud federated learning system, RoadFed, is proposed for road hazard detection. Unlike existing approaches, it uniquely integrates multimodal data from diverse IoT devices to achieve accurate and communication-efficient collaboration without compromising data privacy.}
			\item 
			\textcolor{black}{A new Multimodal Road Hazard Detector (MRHD) is developed. It leverages a triplet loss to capture intra-class and inter-class relationships across multimodal data, achieving superior performance in road hazard detection compared to previous multimodal models.}
			\item
			\textcolor{black}{A novel Federated Multimodal Learning scheme (MFed) is introduced. It significantly reduces communication overhead while ensuring precise detection of road hazards on challenging non-i.i.d. multimodal data.}
			\item 
			\textcolor{black}{An advanced Multimodal Local Differential Privacy algorithm (MLDP) is proposed. It is uniquely designed to mitigate privacy loss caused by the high dimensionality of multimodal data, effectively balancing data privacy and model performance.}
		\end{itemize}
		
		The rest of the paper is summarized as follows. We review the related literature in Section \ref{sec:RelatedWork}. Section \ref{sec:FrameworkDesign} depicts our framework design, including its design goals, framework components, and operational workflow. In Section \ref{sec:Methodologies}, details of the proposed key methodologies are described, consisting of the Multimodal Road Hazard Detector (MRHD), Multimodal Federated Learning Scheme (MFed), and Multimodal Local Differential Privacy Algorithm (MLDP). \textcolor{black}{In Section \ref{sec:Discussion}, we discuss the case of missing modalities, show the latency of RoadFed on typical edge servers, and presented the performance of RoadFed with a different number of clients.} Experimental results are illustrated in Section \ref{sec:Evaluation}, and Section \ref{sec:conclusion} concludes the paper. 
		
		\begin{table*}[t]
			\centering
			\caption{\textcolor{black}{Comparison of RoadFed with existing work in the domain of road hazard detection.}}
			\label{tab:related works}
			\begin{tabular}{ccccccc}
				\toprule
				\textbf{Category} & \textbf{Existing work} &  \textbf{Multi-modal} & \textbf{Fedederated} & \textbf{Privacy-preserving} & \textbf{Non-i.i.d.}\\
				\midrule
				Detectors & \cite{EEFNet,zhao2024community,saha2024federated,moroto2024snow} & $\times$ & $\times$ & $\times$  & $\times$  \\
				& \cite{abbariki2025interpreting,saeed2024multimodal,zhouxiang2025driver,abavisani2020multimodal} & $\checkmark$ & $\times$ & $\times$  & $\times$ \\
				\midrule
				& \cite{Edge-cloudRD,dang2022collaborative,liu2025real} & $\times$ & $\times$ & $\times$ & $\times$  \\
				Distributed systems & \cite{vondikakis2024fedrsc,wu2024hierarchical,dwivedi2024road,saha2024federated} & $\times$  &  $\checkmark$ & $\times$ & $\times$ \\
				& \cite{wu2024hierarchical} & $\times$  &  $\checkmark$ & $\times$ &  $\times$ \\
				& RoadFed (ours) & $\checkmark$  &  $\checkmark$ & $\checkmark$ &  $\checkmark$ \\
				\bottomrule
			\end{tabular}
		\end{table*}
		
		\section{Related Work} \label{sec:RelatedWork}
		IoTs offer promising opportunities to optimize decision-making and improve efficiency. As part of intelligent transportation, many road hazard recognition and alarm techniques using IoT are introduced. Besides, the recent innovations of FL enable collaborative learning. Moreover, different privacy-preservation techniques are utilized to protect privacy. The related literature is summarized as follows.
		
		\subsection{Road hazard/damage detection techniques.}
		\textcolor{black}{Deep learning algorithms are used for the identification of road hazards in many existing studies~\cite{Edge-cloudRD,EEFNet,zhao2024community,saha2024federated,moroto2024snow} and have achieved promising results. The authors of~\cite{EEFNet,zhao2024community,saha2024federated,moroto2024snow} introduced CNN-based models for the classification of road hazards. Despite the success of deep learning models in processing visual data within cluttered real-world scenarios, current road damage detection systems primarily rely on visual input. For example, Wang et al.~\cite{wang2021assistant} used a deep learning algorithm to identify road obstacles, thus mitigating road hazards. In~\cite{sulistyowati2021monitoring}, a threshold-edge-based algorithm was proposed to detect holes in roads and report them on Google Maps.}
		
		\textcolor{black}{Some existing studies \cite{abbariki2025interpreting,saeed2024multimodal,zhouxiang2025driver,abavisani2020multimodal} have also explored the integration of multimodal data in this domain and achieved promising results. 
			In \cite{abbariki2025interpreting}, a multimodal framework was introduced that integrates depth estimation, optical flow, and vision-language models to detect driver reactions to "out-of-label" hazards in autonomous driving scenarios. Saeed et al.~\cite{saeed2024multimodal} developed a multimodal deep learning method that integrates image and audio data to evaluate gravel road conditions. The authors of \cite{zhouxiang2025driver} proposed a multimodal recognition model that utilizes an attention-based intermediate fusion technique to integrate driver video, audio data, and road condition videos for detecting dangerous driving states. The authors of \cite{maeda2025damage} proposed a novel multimodal deep learning model that integrates text data with image analysis to accurately classify damage levels by using an end-to-end attention mechanism that focuses on damaged regions and adjusts its influence based on a confidence score. Abavisani et al.~\cite{abavisani2020multimodal} introduced a multimodal model with a cross-attention module for the categorization of crisis events. Tian et al. \cite{tian2025multimodal} introduced a multimodal deep learning framework that integrates aerial imagery, building footprint data, and traffic flow information to improve traffic risk prediction at urban intersections.
		}
		
		\subsection{Edge-cloud-based distributed monitoring systems.}
		\textcolor{black}{Existing edge-cloud-based distributed monitoring systems can mainly be categorized into road hazard detection-related and task-agnostic. For the former, some studies like \cite{Edge-cloudRD,dang2022collaborative,liu2025real} are edge-cloud-based, but not federated. Specifically, 
			The authors of \cite{Edge-cloudRD} proposed a hybrid edge-cloud framework that leverages lightweight edge AI for real-time anomaly detection and cloud-based generative models for detailed analysis.
			Dang et al. \cite{dang2022collaborative} developed a road damage classification method that uses a standardized entropy threshold to decide whether to process data on an edge device for a fast response or on a cloud server for higher accuracy, with the cloud also assisting in updating the edge model. Liu et al. \cite{liu2025real} proposed a real-time pavement distress detection system based on a lightweight YOLO network (YOLO-LFE), which is designed for deployment on edge devices and reduces parameters and computational requirements compared to the original YOLOv8. Other studies like \cite{EEFNet,vondikakis2024fedrsc,wu2024hierarchical,dwivedi2024road,saha2024federated} applied federated learning for distributed and collaborative road damage detection. Specifically, 
			the authors of \cite{EEFNet} developed an edge-efficient multi-scale focusing diffusion network (EEFNet) that integrates specialized modules, e.g., contour enhancement, multi-scale feature extraction, and dynamic task alignment, for road defect detection on edge devices.
			Vondikakis et al. \cite{vondikakis2024fedrsc} introduced FedRSC, a federated learning system that uses multi-label classification to analyze and identify various road conditions by bringing together edge computing and cloud technology. Wu et al. \cite{wu2024hierarchical} introduced a hierarchical federated learning framework for construction quality defect inspection, which allows robots to collaboratively train a deep learning model. Dwivedi et al. \cite{dwivedi2024road} explored the use of federated learning (FL) for road damage detection across diverse geographical locations, including Japan, China, Norway, and the USA, demonstrating that a collaboratively trained FL model can outperform individual centralized models by leveraging a broader range of data. Saha et al. \cite{saha2024federated} utilized federated learning to develop a global road damage detection model and address limitations of centralized systems.}
		
		\textcolor{black}{Despite the success, most of the existing work in this domain did not consider the high communication cost issue of FL or the low model performance on non-i.i.d. data distributions.
			However, in general domains, extensive research has explored such challenges. Specifically, for reducing communication overhead, the authors of \cite{mcmahan2017communication} achieve this by reducing communication frequency (i.e., each client refreshes its local model multiple times before transmitting it, rather than sending it after every iteration), while \cite{alistarh2017qsgd} accomplishes this by applying model compression techniques (i.e., using model quantization or pruning techniques).
			For achieving high detection accuracy on challenging non-i.i.d. distributions, some researchers \cite{zhao2018federated} introduced a method that generates a slim data pool shared between all edge servers for training, while other work like \cite{li2024iofl,li2024feature}, explored directly learn from non-i.i.d. data. In particular, Li et al.~\cite{li2024iofl} proposed an Intelligent-Optimization-Based Federated Learning (IOFL) framework, where the server directly searches for global model parameters using intelligent optimization algorithms, while clients only validate the model and return test accuracy. This approach fundamentally eliminates the impact of non-i.i.d. data on model performance. The authors of \cite{li2024feature} tackled non-i.i.d. challenges in federated learning by generating privacy-preserving synthetic data that matches essential class-relevant features.}
		
		\begin{figure*}[t]
			\vspace{-2mm}
			\centering
			\includegraphics[width=0.99\textwidth]{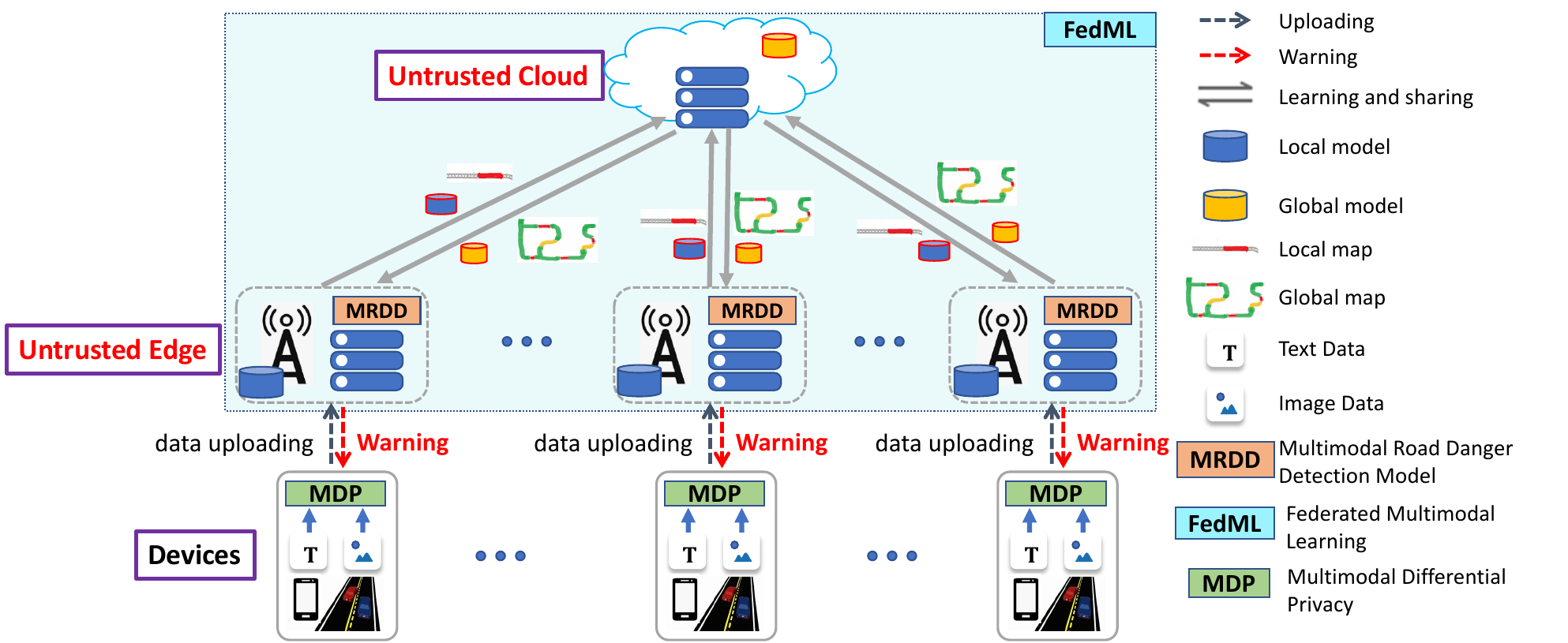}
			\caption{\textcolor{black}{An overview of the proposed RoadFed framework, including three key components (i.e., road users' devices, untrusted edges, and untrusted cloud.) and three key methodologies (i.e., MRHD, MFed, and MLDP).}}
			\label{fi:RoadFedFramework}
			\vspace{-2mm}
		\end{figure*}
		
		\subsection{IoT-based privacy-preserving techniques:}
		\textcolor{black}{Numerous techniques have been put forward to safeguard privacy based on the differential privacy technique.  
			Differential privacy~\cite{dwork2014algorithmic} offers robust privacy assurances that can simultaneously protect user data and model efficacy. LDP is a type of DP that safeguards user data directly from personal devices like smartphones and smartwatches. Consequently, LDP can maintain user privacy without relying on a trusted intermediary (such as unreliable edge or cloud servers). 
			The randomized response method was applied to encode values
			to facilitate local privacy protection. This strategy is straightforward to implement without incurring additional computation costs; however, it performs poorly with high dimensional data. 
			\textcolor{black}{Horigome et al. \cite{horigome2023expectation} proposed an Expectation-Maximization (EM)-based algorithm designed for PrivKV, which focuses on estimating intermediate binary variables instead of continuous values in key-value pairs, thereby making the estimation problem tractable and robust regarding data size and privacy budgets.}
			EM-based methods can cause high variance because of the allocation of the privacy budget, making them less suitable for high dimensional datasets. The authors of~\cite{xia2020distributed} utilized transformation techniques to convert data into binary strings. The randomized response technique was then applied to generate these strings, with the nearest center being communicated differentially privately. 
			Batool et al.~\cite{batool2024secure} introduced a two-layer federated learning framework with local differential privacy at the vehicle level ensures secure and privacy-preserving data sharing in VANETs without relying on trusted third parties.
			Li et al.~\cite{li2025wf} enhances trajectory data utility under local differential privacy by adaptively allocating privacy budgets via water-filling theory and optimizing user segmentation through entropy-driven grouping.}
		
		
		\subsection{Difference between our work and existing research.}
		\textcolor{black}{Our work fundamentally differentiates itself by introducing a holistic and practical solution for road hazard detection that addresses the limitations of existing methods across multiple dimensions. First, while many existing studies on road hazard detection rely on centralized methods, which require a single server to collect and process all data, our approach embraces a distributed, crowdsourced paradigm. This is crucial for addressing the limitations of centralized systems, such as scalability issues and single points of failure, which are particularly relevant in real-world intelligent transportation systems. Second, although some recent works have explored distributed methods for this domain, they often fall short in multimodal data integration, communication efficiency, privacy preservation, or handling non-i.i.d. data (as illustrated in Table \ref{tab:related works}).
		}
		
		\section{Framework Design} \label{sec:FrameworkDesign}
		This section outlines the design objectives, integral parts, and overall working process of the proposed RoadFed framework.
		
		\subsection{Design Objectives}
		The development of RoadFed is guided by the following aims:
		\begin{enumerate}
			\item [--] {\underline{Latency:} The designed system must be capable of detecting road hazards and releasing alarms to road users timely to prevent accidents. Therefore, it is essential to maintain low latency.}
			\item [--] {\underline{Accuracy:} A well-designed road hazard detection system should be able to accurately recognize road hazards, as failing to detect hazards can have serious consequences for road users.}
			\item [--] {\underline{Robustness:} The performance of the developed system should remain stable across various environments, including differing weather and lighting conditions. Additionally, it should maintain high performance even when some edge servers operate with limited and non-i.i.d. data, which is frequently encountered in practical scenarios.}
			\item [--] {\underline{Coverage:} The designed framework should offer extensive coverage to offer users information about hazardous road conditions, facilitating the prevention of road accidents and the planning of safer routes.}
			\item [--] {\underline{Communication and computation overhead:} A good distributed road hazard detection system should have low communication and computation costs for being able to be applied in practical applications.}
			\item [--] {\underline{Privacy:} There is a considerable threat of privacy breaches from untrusted edge servers or clouds, especially during data transmission in open environments. Besides, extensive studies have proved that even only transferring model parameters rather than raw data, attackers can still recover the data utilized for training the model from model parameters, as studied by \cite{geiping2020inverting}. Hence, the designed framework must ensure the protection of user privacy, including personal identifiers and locations, as well as the confidentiality of sensitive information in collected data, such as pictures of person and car plates.}
		\end{enumerate}

		\subsection{Framework Elements}
		The RoadFed framework consists of four essential components, as depicted in Fig.~\ref{fi:RoadFedFramework}.
		\begin{enumerate}
			\item [--] {\underline{Devices:} 
				\textcolor{black}{IoT devices, such as traffic cameras, vehicle sensors, and smartphones, are employed to gather multimodal data (i.e., image-text pairs) and subsequently transfer this information to the adjacent edge server, for example, the Roadside Unit (RSU).}}
			\item [--] {\underline{Edges:} Edge servers are tasked with receiving data from users and swiftly addressing any potential road hazards present in the data. Specifically, the Multimodal Road Hazard Detector (MRHD) is implemented on the edges for the detection of road hazards. The MRHD versions running on the edges and the cloud are referred to as the local and global models. As edges are considered unreliable in this context, it is critical to ensure that sensitive user data from IoT devices remains confidential. Additionally, no data is retained on edge servers, and prior local models are routinely erased to enhance data processing speed.}
			\item [--] {\underline{Cloud:} The cloud functions as an aggregator for FL, facilitating data processing and storage. The global model resides in the cloud server, representing an accumulation of the captured local models. The global map on the cloud server is generated by aggregating the local models and is displayed in real-time on a Google map. This information allows road users to receive timely alerts regarding road hazards and aids in optimizing travel routes. 
				The cloud is also considered unreliable here. The information stored within can be utilized by road management authorities for rapid repairs and effective budget management.}
			\item [--] {\underline{MRHD:} The Multimodal Road Hazard Detector (MRHD) is a deep learning model designed to process both image and text data collected from IoT devices to identify various road hazards, such as significant road damage, collisions involving vehicles, icy conditions, and fallen trees obstructing pathways (refer to Section~\ref{subsec:MRHD}). MRHD is positioned on edge servers to enable prompt detection and alerts concerning road hazards.}
			\item [--] {\underline{MFed:} The proposed Multimodal Federated Learning scheme (MFed) enhances road hazard detection performance through collaborative learning between edges and the cloud server, as edges possess greater computational capabilities and are located more adjacent to users than the cloud server. Many existing federated learning strategies exhibit inefficiencies in communication, so the design of MFed is aimed at significantly reducing communication overhead while guaranteeing high model performance and ensuring fast convergence. Further details regarding MFed are described in Section~\ref{subsec:MFed}.}
			\item [--] {\underline{MLDP:} The developed Multimodal Local Differential Privacy algorithm (MLDP) (refer to Section~\ref{subsec:MLDP}) safeguards both user privacy (such as user identification) and the confidentiality of data collected on users’ devices (e.g., people's faces) before being uploaded to nearby edge servers. MLDP enhances existing local differential privacy algorithms by addressing high expected error rates in high dimensional real-world data. This approach is applied to users’ devices to ensure privacy before sending data to the nearest edge, creating a more secure and user-friendly framework.}
		\end{enumerate}
		
		\begin{figure*}
			\centering
			\includegraphics[width=0.6\textwidth]{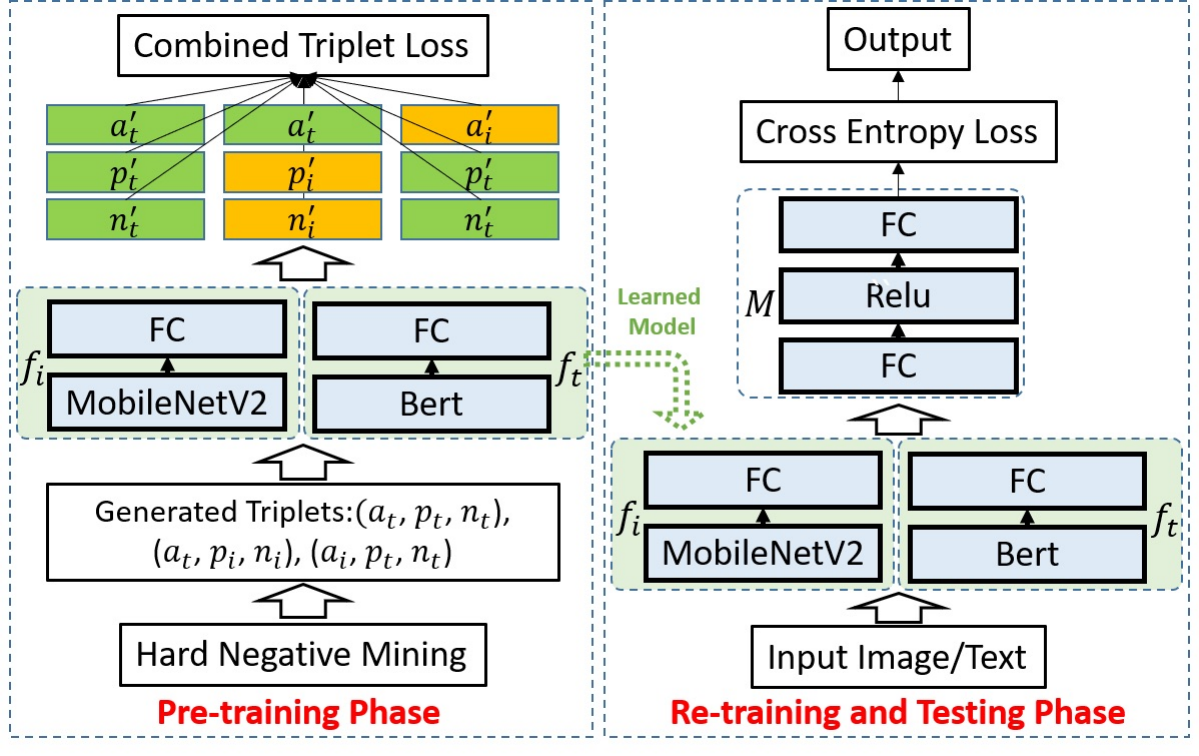}
			\caption{The proposed Multimodal Road Hazard Detector utilizes a triplet loss to improve feature quality, i.e., enlarging inter-class features’ distances and shrinking intra-class features’ distances, for higher accuracy.
			}
			\label{fi:MRHD}
			\vspace{-2mm}
		\end{figure*}
		
		\subsection{Operational Workflow}
		As depicted in Fig.~\ref{fi:RoadFedFramework}, RoadFed is structured on a device-edge-cloud framework where IoT devices facilitate data gathering, an edge server is utilized to minimize response time (i.e., latency), and the cloud server is engaged for aggregation of model parameters. The introduced MRHD is placed at the edge servers for rapid response to road hazards. If a road hazard is identified, the edge server promptly transmits an alarm to road users to prevent road accidents. 
		FL is employed to jointly enhance model training across a number of edges with the help of a cloud. This FL approach allows for effective model development without necessitating the transfer of data from the edges to the clouds, safeguarding data privacy against potentially untrustworthy clouds. In RoadFed, local models refer to those established at the edges, while the road hazard detector in the cloud server is referred to as the global model. The operational workflow of RoadFed consists of four key phases that continuously learn from edge data.
		
		\begin{itemize}
			\item \textbf{Stage 1:} Each road user gathers image or textual information regarding road hazards using smart IoT devices and subsequently transmits this data to the nearby edge server.
			\item \textbf{Stage 2:} Edges assess road hazards within their communication vicinity utilizing the MRHD model (received from the cloud). Following this, they disseminate road hazard alarms to all road users within their coverage area. Edge servers initiate the training of their local models on their local datasets when the newly accumulated data surpasses a predetermined threshold (configured to 100 based on comprehensive testing). Subsequently, they transmit the updated local models' parameters to the cloud server.  
			The local models on the edge servers that do not have sufficient new data will not be trained to minimize communication and computation costs.
			\item \textbf{Stage 3:}
			The cloud server integrates the local parameters obtained from the covered edges according to Eq.~\eqref{eq:Fed} and Eq. (2) to formulate a global model.
			\begin{align}\label{eq:Fed}
				\boldsymbol{\omega}^t &= \sum_{i=1}^{N} \frac{D_i}{D} \boldsymbol{\omega}_i^t, \\
				D &= \sum_{i=1}^{N} D_i,
			\end{align}
			in which $\boldsymbol{\omega}^t$ signifies the weights of the global model at time $t$, and $\boldsymbol{\omega}_i^t$ indicates the weights of the $i$-th local model at time $t$. $D_i$ represents the size of the training dataset of the $i$-th edge, while $D$ is the size of the overall training dataset across all participating edge servers. Here, $N$ denotes the count of edges that have transmitted their local models to the cloud. Subsequently, the cloud sends the global model to all edges within its coverage.
			\item \textbf{Stage 4:}
			Edge servers update their local models' parameters using the global model's parameters received from the cloud server.
		\end{itemize}
		Stages 1 to 4 are reiterated in every $R$ communication round.
		
		\textcolor{black}{Please note that the assumption of untrusted edge and cloud servers primarily focuses on mitigating privacy risks, as they could potentially misuse or leak sensitive user data. \textcolor{black}{This is a common assumption in many federated learning research works like \cite{EEFNet,hussain2024ensuring} to address the worst-case privacy threat.} The proposed MLDP and MFed mechanisms are specifically designed to address this privacy concern.}

		\textcolor{black}{\textbf{Computational complexity analysis.} 
			For any participant of a RoadFed system, its overall computation complexity is determined by the most complex component, i.e., MRHD. The computational complexity of MFed and MLDP is negligible compared with MRHD. MRHD has $3.2\times10^{10}$ FLOPs and $1.1\times10^8$ parameters, which define the model complexity of RoadFed.
		}
		
		\section{Methodologies} \label{sec:Methodologies}
		This section describes the details of the proposed Multimodal Road Hazard Detector (MRHD), Multimodal Federated Learning Scheme (MFed), and Multimodal Local Differential Privacy Algorithm (MLDP).
		
		\subsection{Multimodal Road Hazard Detector} \label{subsec:MRHD}
		\textcolor{black}{The architecture of the proposed Multimodal Road Hazard Detector (MRHD) encompasses two stages: the pre-training stage and the re-training and testing stage, as depicted in Fig.~\ref{fi:MRHD}. 
			During the pre-training stage, the model is pre-trained using a specially designed triplet loss to capture both intra-class and inter-class relationships within multimodal data. In the second stage, the pre-trained multimodal feature extractor is applied to the road hazard classification task. 
			The novelty of MRHD lies in the special design of pretraining the model using triplet loss so that the model's performance can be enhanced on tasks with limited sample sizes.} 

The goal of the pre-training stage is to train both text and image feature extractors (i.e., $f_t$ (Bert+FC) and $f_i$ (MobileNetV2+FC)) so that they can differentiate between benign road conditions and various kinds of road hazards from text/image data, where FC denotes a fully connected layer. Cosine similarity is employed to evaluate the distances between embeddings.

A designed triplet loss is formulated in Eq.~\eqref{eq:combinedTripletLoss}, which measures the intra-class and inter-class relationships of different data modalities, where $\alpha$ represents a penalty factor that regulates the significance of the term. The designed triplet loss comprises fundamental triplet losses for text-only Eq.~\eqref{eq:tripletLoss1}, text-image Eq.~\eqref{eq:tripletLoss2}, and image-text Eq.~\eqref{eq:tripletLoss3}. For the experiments, we set $c=0.2$ and $m=0$ across all trials.

\begin{equation} \label{eq:combinedTripletLoss}
	Loss = \alpha \cdot Loss(a_t, p_t, n_t) + Loss(a_t, p_i, n_i) + Loss(a_i, p_t, n_t), 
\end{equation}
\begin{equation} \label{eq:tripletLoss1}
	Loss(a_t, p_t, n_t) = max\{\cos(a_t, p_t) - \cos(a_t, n_t) + c, m\}, 
\end{equation}
\begin{equation} \label{eq:tripletLoss2}    
	Loss(a_t, p_i, n_i) = max\{\cos(a_t, p_i) - \cos(a_t, n_i) + c, m\},
\end{equation}
\begin{equation} \label{eq:tripletLoss3}    
	Loss(a_i, p_t, n_t) = max\{\cos(a_i, p_t) - \cos(a_i, n_t) + c, m\}.
\end{equation}

Training the model with the designed triplet loss function over a large number of triplets can be computationally intensive. 
We select the most violating negative data points within every batch. 
In particular, feature vectors of three triplets, namely $(a_t, p_t, n_i)$, $(a_t, p_t, n_t)$, and $(a_i, p_t, n_t)$ are chosen in every batch, ensuring that the most difficult negative sample is employed for training in every batch; here, $a$, $n$, and $p$ denote anchor, negative, and positive data points, respectively, while $t, it$ refer to text and image data modalities. A negative sample is selected if the cosine similarity among an anchor sample and its negative pair is smaller than the cosine similarity of it to any other negative samples within the batch.

In the second stage, the MRHD is built based on the pre-trained image and text feature extractors (i.e., $f_i$ and $f_t$) as well as a merging block $m$. 
The cross-entropy loss is utilized to optimize the detector.
The merging block $m$ consists of two FC layers and one ReLU layer, as depicted in Fig. \ref{fi:MRHD}.

\textcolor{black}{
	The multimodality feature fusion in the MRHD model is a late fusion because, as shown in Fig.~\ref{fi:MRHD}, the MRHD algorithm takes image-text pairs as input, where image features are first independently extracted using MobileNetV2 and text features are independently extracted using BERT. These high-level features from each modality are then concatenated and passed through two fully connected (FC) layers to produce the final output. This process aligns with the core characteristics of late fusion: each modality is processed separately to obtain its own high-level representations before fusion occurs at a later stage, rather than fusing raw inputs or low-level features early on.
}

\textcolor{black}{The proposed Multimodal Road Hazard Detector (MRHD) identifies road hazards using images-text pairs as inputs, but when encountering the case of missing modalities, with a simple masking mechanism, MRHD can operates without the need for paired image-text data. More details can be found in Section \ref{sec:Discussion}.}

\subsection{Multimodal Federated Learning Scheme} \label{subsec:MFed}
\textcolor{black}{The Multimodal Federated Learning Scheme (MFed) is designed to obtain superior detection performance across various edge servers with minimal communication overhead while ensuring the convergence of the model on non-i.i.d. datasets. MFed is primarily composed of three components: periodic parameter sharing, adaptive learning rate, and dynamic quantization. Periodic parameter sharing refers to the edge servers transmitting their trained local models after $E$ local training epochs instead of sharing them after every local training epoch to reduce the total number of transmissions between edge servers and the cloud server. The adaptive learning rate strategy means that the learning rate of the models is decayed in a specific way during FL training to ensure fast convergence on non-i.i.d. data. The dynamic quantization technique directly reduces the model size by quantizing model parameters. 
	The novelty of MFed lies in the integration of simple but efficient techniques to provide an effective solution for accurate and communication-efficient road hazard detection in distributed systems operating on real-world, non-i.i.d. multimodal datasets. 
	The followings are detailed design of the adaptive learning rate and dynamic quantization strategy.}

\subsubsection{Adaptive Learning Rate}
Although existing FL strategies~\cite{mcmahan2017communication,reisizadeh2020fedpaq,li2019convergence} have achieved promising results, there are still some open issues that need to be solved, for example, high convergence time, particularly with challenging non-i.i.d. data. To address this challenge, the learning rate (LR) in MFed is reduced according to Eq.~\eqref{eq:AdaLR} after each global round, following~\cite{li2019convergence}.
\begin{equation} \label{eq:AdaLR}
	\gamma_r = \gamma_0 \cdot \delta^{\floor*{\frac{\nu}{\zeta}}},
\end{equation}
where $\gamma_0$ represents the initial LR, set to 0.1 for the experiments. $\delta$ is set to 0.5. $\zeta$ and $\nu$ denote step size and the last global round, while $\zeta$ is configured to 1. Reducing the learning rate is essential to ensure the global model's convergence when working on non-i.i.d. datasets~\cite{li2019convergence}. 

\subsubsection{Dynamic Quantization}
One main challenge of FL is the substantial bandwidth costs incurred from constant parameter communications between the cloud server and edge servers. In MFed, at time $y$, each participating edge communicates only the quantized parameter differences ${\Delta \omega}_i^t$ between the obtained global model $ \boldsymbol{\omega}^{t-1}$ at time $t-1$ and the newly trained local model $\boldsymbol{\omega}_{i}^t$ at time $y$ to the cloud server, rather than transmitting the entire local model $\boldsymbol{\omega}_{i}^t$. The Low-Precision Quantizer (LPQ)~\cite{alistarh2017qsgd}, specifically the QSGD method, is employed to compress these model differences, as it provides convergence guarantees along with strong practical performance. The trade-off between convergence time and communication overhead can be adjusted smoothly (i.e., on a per-iteration basis) using QSGD.

After the local models are trained and aggregated, or before the local or global models are sent, the dynamic quantization technique\footnote{https://pytorch.org/tutorials/recipes/recipes/dynamic\_quantization.html} is employed to further diminish the model size and improve its efficiency by simply converting float32 into int8 values.
Besides, due to the precise calculation of the signal range for each input, it can substantially reduce latency without compromising accuracy significantly \cite{Gholami2022ASO, DQ}.
The primary concept behind it is to adaptively decide the degree of compression, ensuring that the most critical information is preserved while keeping a low model size. The proposed MFed strategy is formally outlined in Algorithm~\ref{alg:MFed}.

\begin{algorithm}[t]
	\caption{MFed scheme}\label{alg:MFed}
	\SetKwInOut{Input}{Input}  
	\SetKwInOut{Output}{Output}
	\Input{Datasets at edge servers and the detector MRHD}
	\Output{Trained local models at edge servers}
	The cloud server initialize $\boldsymbol{\omega}^0$ and distributes it to the covered edges \\
	The cloud server sets the initial LR as $\gamma_0$ \\
	{ \For{each global communication round $R$}{
			\For{each E epochs}{
				\For{every edge $i \in \{1,2,\cdots,K\}$}
				{
					Each edge server replaces its local model $\boldsymbol{\omega}_{i}^t$ with the obtained global one $\boldsymbol{\omega}^{t-1}$ \\
					Each edge server trains its local model $\boldsymbol{\omega}_{i}^t$ on its newly acquired local data by performing:
					$\boldsymbol{\omega}_{i}^t \xleftarrow{} \boldsymbol{\omega}_{i}^{t-1} - \frac{\gamma_0}{R+1} \bigtriangledown l(\boldsymbol{\omega}_{i}^{t-1}, b_{i}^{t-1})$ \\
					Computes the weight difference by:
					${\Delta \omega}_i^t = \boldsymbol{\omega}_{i}^t - \boldsymbol{\omega}^{t-1}$ \\
					Each edge server applies the dynamic quantization technique on the $Q({\Delta \omega}_i^t)$ \\
					Each edge server transmits $Q({\Delta \omega}_i^t)$ to the cloud server \\
					$R = R+1$ \\
			}}
			The cloud server waits until $K$ local models are gathered \\
			The cloud server integrates the local models by: $\boldsymbol{\omega}^t = \boldsymbol{\omega}^{t-1} + \frac{1}{K} \sum_{i=1}^{K} Q({\Delta \omega}_i^t)$ \\
			The cloud server applies the dynamic quantization method on the global model $\boldsymbol{\omega}^t$ \\
			The cloud server transmits the global weights $\boldsymbol{\omega}^t$ to the covered edges servers  \\
	}}
\end{algorithm}

\subsection{Multimodal Local Differential Privacy Algorithm} \label{subsec:MLDP}
The Multimodal Local Differential Privacy algorithm (MLDP) seeks to decrease any detrimental impact on MRHD's performance while effectively protecting private information. MLDP is designed following the Local Differential Privacy (LDP) proposed by \cite{kairouz2014extremal} that is implemented on IoT devices to alter data prior to transmission to potentially unreliable edge servers. 
\textcolor{black}{Different from existing LDP techniques like \cite{kairouz2014extremal}, MLDP employs a specialized mapping method for feature dimensionality reduction, which is designed to better preserve important information of the original input while largely reducing the input dimensionality (thereby decreasing the amount of introduced noise).}

When a user collects a text $y$ or image $x$, the Laplace Mechanism is utilized to introduce perturbations, which is one standard distribution of $\epsilon$-LDP. Specifically, the perturbed text or image data $X$ can be denoted as follows:
\begin{equation} \label{eq:addNoise1}
	\forall j \in [d], X^*[j] = X[j] + Laplace(\frac{s_1(f)}{\epsilon}),
\end{equation}
where $Laplace(\frac{s_1(f)}{\epsilon})$ means a Laplace distribution with scale $\frac{2d}{\epsilon}$.
The error derived from perturbing the input samples using the LDP Algorithm is $O(\frac{d}{\epsilon})$, where $d$ is the dimension of the input data. It could be extremely high for high dimensional data.
To mitigate the problem, we intentionally decrease the dimension of the data before applying the LDP.

As stated by~\cite{achlioptas2001database}, mapping a vector into a randomly selected lower-dimensional subspace can still capture important characteristics. However, this method is limited to reducing dimensions by a factor of up to $\sqrt{d}$, which may still be substantial when $d$ is big. To address this limitation, the dimensionality is further reduced by mapping the input to a smaller subset, ensuring that important information is preserved. Specifically, text data is first encoded into numerical vectors. The dimensions of both image and text data are then reduced by multiplying by random matrices $Q_{c \times d}$ ($c < d$) and $R_{d \times e}$ ($e < d$), generated \textcolor{black}{by the devices}. Each element of $Q$ and $R$, namely, $Q[i][j]$ and $R[i][j]$, is denoted as follows:
\begin{equation} \label{eq:dimensionReduction}
	Q[i][j] =  R[i][j] = sign(x) \times \frac{1}{e},
\end{equation}
where $x$ is evenly selected from $U(-1, 1)$. $e$ represents the output's dimensionality. Consequently, the altered text is $T = Tanh(Q \times T)$ while the modified image is $I = Tanh(Q \times I \times R)$.

The concept of $\epsilon$-LDP is presented as follows, following~\cite{dwork2014algorithmic}:
\textbf{Definition 1} ($\epsilon-LDP$). \textit{A randomized function $f$ achieves  $\epsilon-LDP$ only if for any two inputs $x$ and $y$, where $\epsilon>0$, it holds that}
\begin{equation}
	P[f(\chi)=\chi^*] \leq exp(\epsilon) \cdot P[f(\chi')=\chi^*], 
\end{equation}
where $P[\cdot]$ means probability, and $\epsilon$ represents the privacy budget, which quantifies the level of noise introduced into the dataset. 
A lower $\epsilon$ means a higher amount of added noise, resulting in enhanced privacy but correspondingly reduced accuracy. 
Based on this definition, the edge servers that capture the altered data $\chi^*$ cannot confidently discover the true value of $\chi^*$ (governed by $\epsilon$), no matter the amount of knowledge the edge servers possess.

To ensure that the data is $\epsilon$-LDP private, a random noise sampled from the Laplace distribution $Laplace(\frac{s_1(f)}{\epsilon})$ is added to the data.
The sensitivity estimates the maximum difference in output that can occur due to noise addition while still preserving privacy. The $L1$-sensitivity is denoted as follows:
\begin{equation}
	\label{eq:RoadFedupperBound}
	s_1(f)= max\{ \lVert f(\chi)-f(\chi') \rVert_{1} \},
\end{equation}
where $\lVert.\rVert_{1}$ refers to the L1 norm.

In this context, $f$ complies with $\epsilon-LDP$.

\textit{Proof:} 
Let $x$ and $y$ represent two samples, each of dimensionality $d$, another independent data point be $x$ (also with dimension $d$), and $d$ random variables are from $Laplace(0, \frac{s_1(f)}{\epsilon})$. 
\begin{align}
	& \frac{Pr[(f(\chi)=\chi^*]}{Pr[(f(\chi')=\chi^*]}
	=\prod_{i=1}^d \frac{exp(-\frac{{\epsilon}|f(\chi)_i-\chi^*_i|}{s_1(f)})}{exp(-\frac{{\epsilon}|f(\chi')_i-\chi^*_i|}{s_1(f)})}, \nonumber \\
	&\quad=\prod_{i=1}^d exp(\frac{\epsilon(f(\chi')_i-\chi^*_i|-|f(\chi)_i-\chi^*_i|)}{s_1(f)}),  \nonumber \\ 
	& \quad\leq \prod_{i=1}^d
	exp(\frac{\epsilon|f(\chi)_i-f(\chi')_i|}{s_1(f)}), \nonumber \\
	&\quad=exp(\frac{\epsilon \lVert f(\chi)-f(\chi') \rVert_{1}}{s_1(f)}), \nonumber \\
	&\quad\leq exp(\epsilon).
\end{align}
Therefore,
\begin{equation} \label{eq:proveFinal}
	Pr[f(\chi)=\chi^*] \leq exp(\epsilon) \cdot Pr[f(\chi')=\chi^*].
\end{equation}

\textcolor{black}{The proof shows that MLDP provides a quantifiable and theoretically sound privacy guarantee.}
Importantly, post-processing invariance is a fundamental property of differential privacy. All computations performed on the edges using data received from IoT devices remain within the bounds of $\epsilon-LDP$. The specifics of the MLDP approach are outlined in Algorithm~\ref{alm:MLDP}.
\textcolor{black}{Please note that the definition of LDP, L1-sensitivity, and the proof are based on established principles from \cite{dwork2014algorithmic}.}

\begin{algorithm}[t]
	\SetKwInOut{Input}{Input}
	\SetKwInOut{Output}{Output}
	\Input{High-dimensional multimodal data (i.e., text $y$ and image $x$) with dimension $d$, and privacy budget $\epsilon$}
	\Output{Privacy-preserved mutimodal data features (i.e., text feature $y^{\prime\prime}$ and image feature $x^{\prime\prime}$)}   
	Create random matrices $Q_{c \times d}$ and $R_{d \times e}$ where each element has an equal chance of being $1/e$ or $-1/e$ \\
	Cut down the dimension of $x$ or $y$ by \\
	$y^\prime_{c \times 1} = Tanh(Q_{c\times d} \times y_{d \times 1})$  \\
	$x^\prime_{c \times e} = Tanh(Q_{c\times d} \times x_{d \times d} \times R_{d \times e}) $  $\xleftarrow{}$ only if the dimension of the text is large \\
	\For {$j = 1, 2, \cdots, d$}
	{
		$y^{\prime\prime}[j] = y^\prime[j] + Laplace(\frac{s_1(f)}{\epsilon})$ \\
		$x^{\prime\prime}[j] = x^\prime[j] + Laplace(\frac{s_1(f)}{\epsilon})$ \\
		Return $ y^{\prime\prime}, x^{\prime\prime}$
	}
	\caption{MLDP}
	\label{alm:MLDP}
\end{algorithm}

\begin{figure}[!ht]
	\vspace{-2mm}
	\centering
	\includegraphics[width=0.4\textwidth]{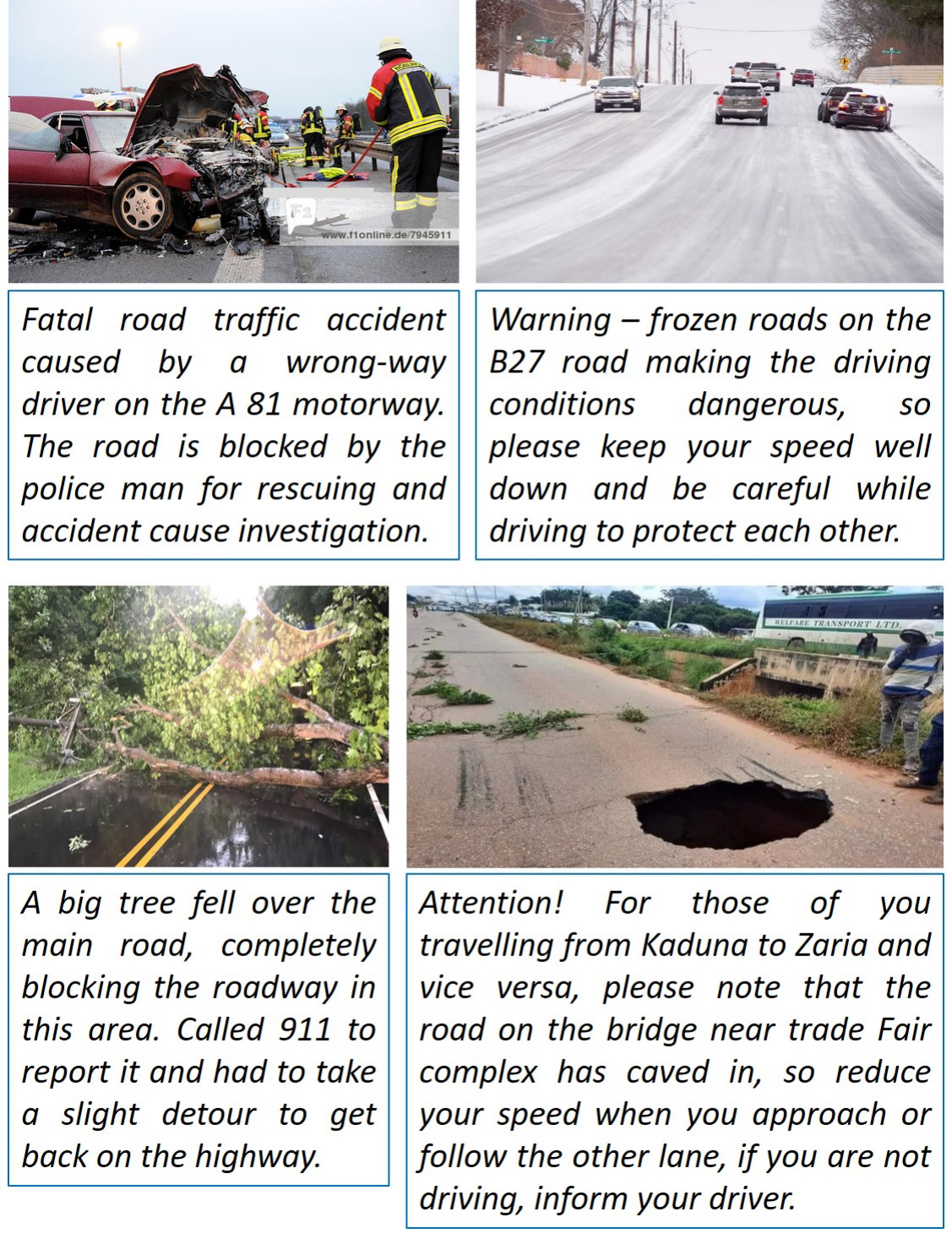}
	\caption{Example images and texts of road hazards (from left to right, crashed vehicles, icy road, fallen tree, and damaged road), including dangerous type and location.}
	\label{fi:dataExample}
	\vspace{-4mm}
\end{figure}

\begin{table*}[!ht]
	\centering
	\caption{\textcolor{black}{Detailed description of the datasets used in the experiments.}} \label{tab:dataset}
	\begin{tabular}{ccccc}
		\toprule
		Dataset & Classes & Total number of samples & Train & Test \\
		\midrule
		\multirow{5}{*}{CrisisMMD (Humanitarian task)} 
		& Infrastructure and utility damage  &  673 & 595 & 78 \\
		& Vehicle damage& 20 & 17 & 3 \\
		& Rescue volunteering or donation effort & 1038 & 912 & 126  \\
		& Other relevant information & 1514 & 1279 & 235  \\
		& Affected people  & 80 & 71 & 9 \\
		\midrule
		\multirow{1}{*}{MNIST} 
		& 10 digits  & 7000$\times$10 & 6000$\times$10 & 1000$\times$10  \\
		\midrule
		\multirow{5}{*}{Our dataset} 
		& Normal  & 487 & 397 & 90 \\
		& Crashed vehicle  & 211  & 167 & 44 \\
		& Damaged road  & 641 & 513 & 128  \\
		&  Fallen tree & 217 & 173 & 44\\
		&  Icy road & 188 & 158 & 30\\
		\bottomrule
	\end{tabular}
\end{table*}

\begin{table}[!ht]
	\caption{\textcolor{black}{Experimental parameter setup.}}
	\label{tab:parameterSetup}
	\center
	\setlength\tabcolsep{1pt}
	\begin{tabular}{ccc}
		\toprule
		Parameter & Our dataset  &  MNIST dataset  \\
		\midrule
		Non-i.i.d. & sampling by class & sampling by shards\\
		setting I & 4 classes per client & [195, 642, 363] shards\\
		&   & for clients [1,2,3] \\
		Non-i.i.d. & Direchlet distribution &Direchlet distribution\\
		setting II & $\beta=[0.2,0.3,0.4,0.5]$ & $\beta=[0.2,0.3,0.4,0.5]$\\
		CC & 50 & 200\\
		Local epoch & 10 & 10 \\
		Number of clients & [3, 8, 15] & [3, 8, 15] \\
		Local batchsize & 16 & 512\\
		Initial learning rate & 0.01 &0.001\\
		\bottomrule
	\end{tabular}
	\vspace{-1mm}
\end{table}

\section{Evaluation} \label{sec:Evaluation}
\subsection{Experimental Setup} 
\subsubsection{Configuration and Implementation Details}
\begin{figure}
\centering
\includegraphics[width=0.7\linewidth]{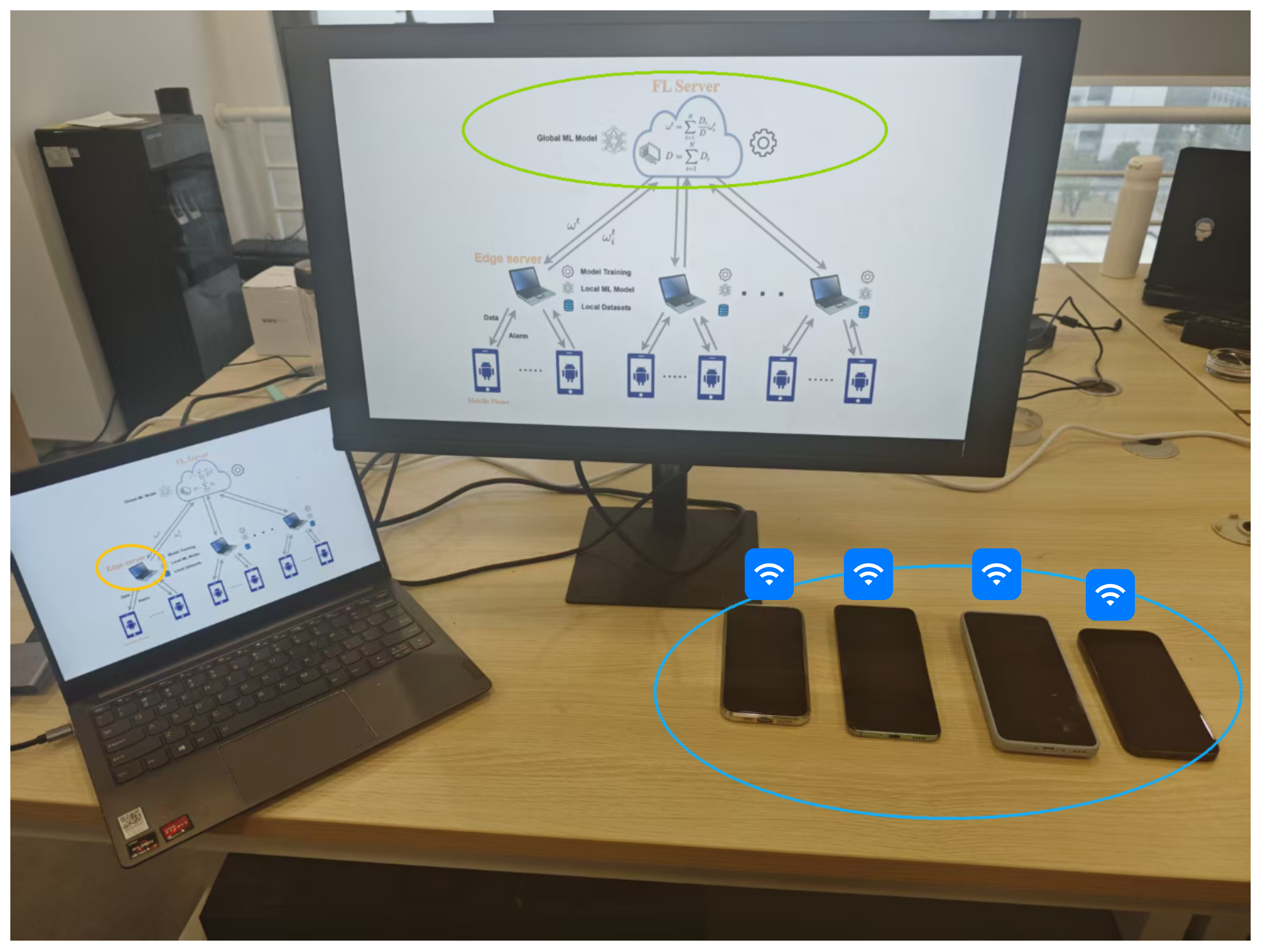}
\caption{\textcolor{black}{Experimental Configuration.}}
\label{fig:implementation}
\end{figure}

\textcolor{black}{The RoadFed framework is based on a device-edge-cloud architecture, with the cloud and edge servers represented by a server (running Ubuntu 22.04, equipped with 384 GB of RAM, AMD Epyc 7282, 16 cores 32 threads and a RTX 3090 GPU) and several high-performance laptops (64-bit Windows 11 with 32 GB RAM, intel Core i9-14900HX, 24 cores (8P+16E) 32 threads, and a RTX 3090 GPU), and a set of budget-friendly smartphones are used to simulate IoT devices, as illustrated in Fig. \ref{fig:implementation}. The configuration of edge servers aligns with the benchmark platforms adopted in recent edge computing research \cite{cai2024edge,huang2025llms}. Except for the Laptop, we also evaluate the model's inference latency on other edge devices (e.g., Jetson AGX Orin and Jetson Orin NX) to verify its practicality under typical edge servers. 
}

\textcolor{black}{The system is developed using the Anaconda environment, with pre-installed Python 3.9.21, PyTorch (version 2.10.0+cu128), and Flower (version 1.26.1). The entire framework is constructed around three Python scripts: one dedicated to cloud server operation, another for edge server operation, and a third for defining specialized model training methodologies. The inter-device communication is carried out through the local area network (LAN), to which all hardware nodes are connected via Wi-Fi. Data transmission and reception between various devices are realized through the Flower framework and gRPC protocol, supported by the high-speed Realtek RTL8125 2.5GbE wired network adapter.}

\textcolor{black}{In the following, we describe in detail the configurations used in the simulations. IoT devices, serving as initial data sources, actively push their local datasets to the edge servers using the standard File Transfer Protocol (FTP) over Wi-Fi. This data transfer process precedes or runs in parallel with the federated learning rounds, ensuring edge servers possess fresh and relevant data for local training. Edge servers (i.e., laptops) establish a gRPC connection to the central FL server (typically at a specified IP address and port, e.g., "10.90.0.8:8080") via the Flower client library. The cloud server dictates the start of FL training rounds. It dispatches the current global model parameters, optionally collects initial parameters if required, and sends a configuration dictionary to the edge servers. This data is serialized into a binary stream and transmitted over the network via gRPC. Upon receiving parameters and configuration, each edge server loads the global model parameters into its local model and performs local training using its private dataset. Following local training, the edge client retrieves its updated local model parameters, its dataset size, and any relevant client-side metrics. These results are encapsulated by the Flower library and transmitted back to the server as a gRPC response.}

\textcolor{black}{Table \ref{tab:parameterSetup} presents a detailed experimental setup for our self-collected and the MNIST datasets.}

\subsubsection{Datasets}
To validate the effectiveness of the proposed algorithms, we conduct experiments on three datasets: CrisisMMD (Humanitarian task)~\cite{alam2018crisismmd} (hereinafter referred to as CrisisMMD), MNIST, and a dataset collected by ourselves (hereinafter referred to as our dataset or our self-collected dataset). 
The CrisisMMD dataset contains five categories, namely infrastructure and utility damage, vehicle damage, rescue volunteering or donation effort, other relevant information, and affected people (including affected individuals/injured or dead people/missing or found people). The MNIST dataset contains 10 classes of handwritten digits. 

\textcolor{black}{Our dataset consists of images paired with corresponding textual descriptions from five categories: normal road (487 samples), crashed vehicle (211), damaged road (641), fallen tree (217), and icy road (188). The text-image pairs consist of images paired with short descriptions (e.g., "A big tree fell over the main road, completely blocking the roadway on Broadway. Called 911 to report it and had to take a slight detour to get back on the highway.") sourced from platforms like social media platforms (e.g., Facebook and Twitter). For pairs with short and relevant text, the data is saved directly. If the text is too long, the title or keywords are selected instead. For images without accompanying text, manual annotation is performed to generate descriptive captions. This simulates a real-world scenario where users report hazards via an app with both a photo and a brief text description. The collected image-text pairs are utilized for simulating data streams generated by user devices in real-world environments. These data are divided into training and testing sets in an 80:20 ratio after desensitization (e.g., removing names and replacing location information) and formatting (e.g., deleting redundant text and checking that the image matches the text description).
} 

\textcolor{black}{A detailed description of the datasets is provided in Table~\ref{tab:dataset}. Fig. \ref{fi:dataExample} displays representative examples of the collected hazards (e.g., crashed vehicles, icy roads, fallen tree, and damaged road) alongside their textual descriptions and location data.}

\textbf{Non-i.i.d. distribution.} \textcolor{black}{To validate the performance of RoadFed under different non-i.i.d. distributions, we simulated two different non-i.i.d. scenarios. For the first non-i.i.d. distribution, we randomly select samples from 1 to 5 distinct classes from our dataset. In the case of "each client has 2 classes," each client possesses samples from any two randomly selected classes in the dataset, and federated learning is conducted among clients based on this configuration. For the MNIST dataset, the 60,000 training images is first randomly partitioned into 1,200 shards, with each shard containing 50 images. This process creates a highly non-i.i.d. data distribution, as each shard is likely to contain data from only a few specific classes. Next, a random number of shards, is assigned to each participant client. For the second non-i.i.d. distribution, similar to \cite{li2022federated}, the data is partitioned by sampling from the Dirichlet distribution with different concentration parameter $\beta$ (a smaller $\beta$ value leads to a higher degree of non-i.i.d.) to assign different class proportions to each client, simulating realistic data heterogeneity scenarios. The Dirichlet distribution is well-suited for modeling the heterogeneous data distributions encountered in real-world federated learning scenarios \cite{li2022federated}. These allocation strategies guarantee a high degree of real-world data heterogeneity among clients.}

\textbf{Data preprocessing.} Prior to the training of the model, images are clipped into $256 \times 256$ for subsequent processing to reduce the training time. Similarly, the texts are tokenized at the word level. Because the text data $y$ only has one dimension and it is relatively small (i.e., 32) for our dataset, the dimensionality of $y$ remains unchanged. 
\textcolor{black}{
	The proposed MRHD algorithm is assessed using both the CrisisMMD and our datasets. The MNIST and our datasets are utilized to evaluate the performance of the introduced MFed strategy and draw comparisons with leading FL strategies.
}

\subsubsection{Baselines}
\textcolor{black}{We compare the proposed MFed strategy with existing ones, like FedAvg \cite{mcmahan2017communication}, FedPAQ \cite{reisizadeh2020fedpaq}, LRDevay \cite{li2019convergence}, FedProx \cite{li2020federated}, and SCAFFOLD \cite{karimireddy2020scaffold}. FedAvg is a federated learning framework that works by having a central server average the model weights received from participating clients after they've performed local model training on their own private data. FedPAQ is a communication-efficient federated learning method that reduces communication overhead and improves scalability by using periodic model averaging, partial device participation, and quantized message-passing. LRDevay provides a theoretical analysis of the FedAvg algorithm's convergence and highlights the necessity of a decaying learning rate to speed up FedAvg's convergence. 
	FedProx is a robust federated learning framework that tackles system and statistical heterogeneity by introducing a proximal term to constrain local updates, mitigating client drift caused by non-IID data and system variability. 
	SCAFFOLD is a communication-efficient federated learning algorithm that corrects client drift caused by data heterogeneity through control variates that reduce variance in local updates. 
	MFed-Q is the proposed MFed strategy without using quantization for reducing model size. MFed-LRD is the proposed MFed strategy without decaying the learning rate.}

\textcolor{black}{In addition, the developed RoadFed is also compared with a series of representative baseline models, including 
	Edge-cloudRD \cite{Edge-cloudRD}, EEFNet \cite{EEFNet}, Vondikakis et al. \cite{vondikakis2024fedrsc},  Dwivedi et al. \cite{dwivedi2024road}, Saha et al.~\cite{saha2024federated}, and Wu et al. \cite{wu2024hierarchical}, to evaluate the effectiveness of our proposed method. 
	Edge-cloudRD is a hybrid edge-cloud framework that leverages lightweight edge AI (MobileNet) for real-time anomaly detection and cloud-based generative models for detailed analysis.
	EEFNet is an edge-deployed road defect detection method that utilizes an enhanced contour extraction module, multi-scale convolution, a context diffusion pyramid network, and a dynamic task-aligned detection head for accurate road defect detection.
	Vondikakis et al. \cite{vondikakis2024fedrsc} is a federated learning system that identify various road conditions by bringing together edge computing and cloud technology. Dwivedi et al. \cite{dwivedi2024road} used federated learning (FL) for road damage detection across diverse geographical locations. Saha et al.~\cite{saha2024federated} applied federated learning with a CNN model for road damage detection.
	Wu et al. \cite{wu2024hierarchical} is a hierarchical federated learning framework for construction quality defect inspection, which allows robots to collaboratively train a deep learning model.
}

\subsubsection{Evaluation metrics}
\begin{figure}[!ht]
\centering
\includegraphics[width=0.4\textwidth]{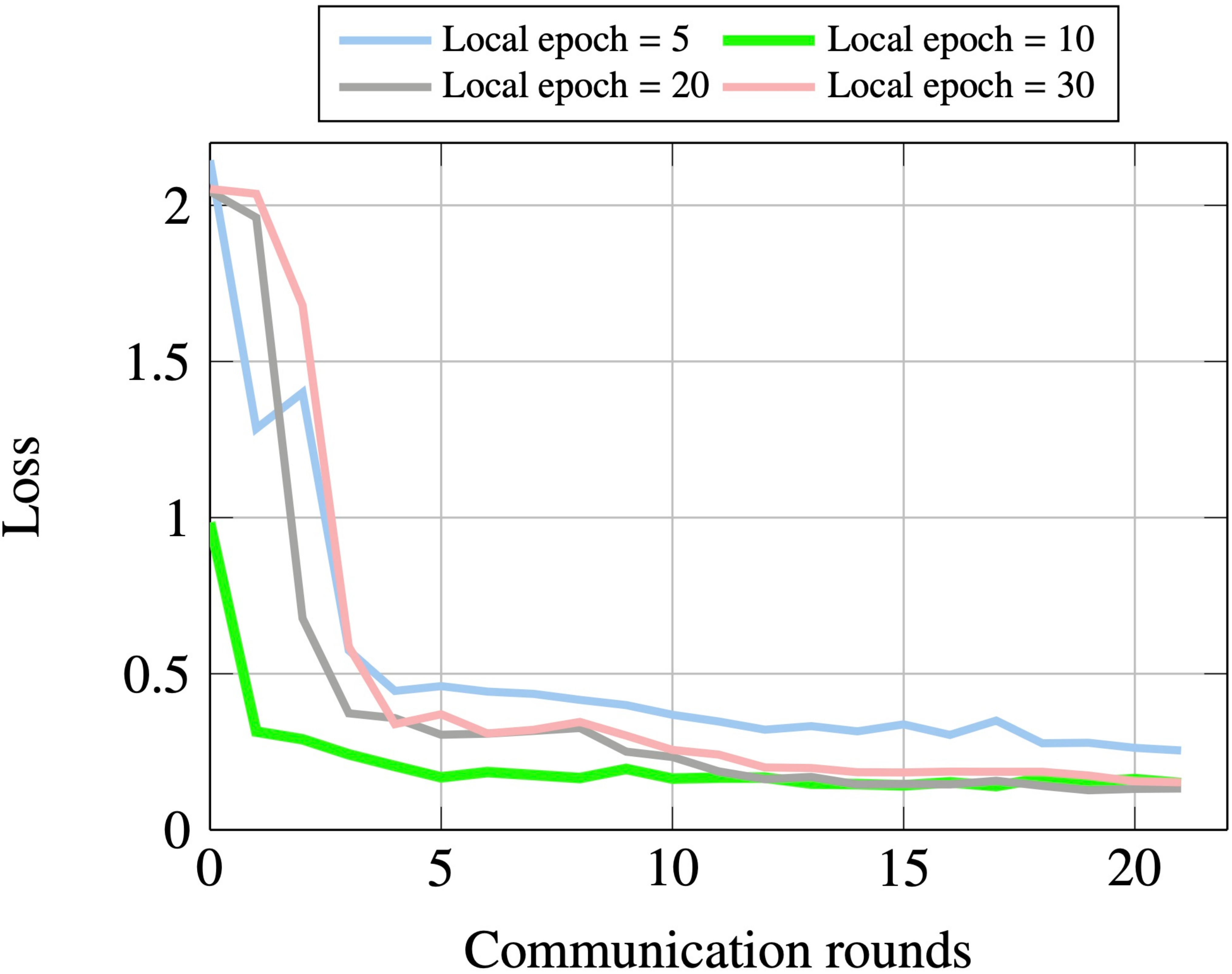}
\caption{\textcolor{black}{Loss of MFed under different local training epochs on our dataset using the MRHD model (each client has 4 classes).} }
\label{fig:FL_diff_localEpoch}
\end{figure}

We use metrics like Accuracy (Acc), Precision, Recall,  F1-score (F1), Latency, Communication Cost (CC), Collaborative Learning (CL), Multi-Modal learning (MM), and Privacy-Preserving (PP) to compare model performance. 
\textcolor{black}{
	Latency refers to the waiting time for a driver to receive a hazard warning, which occurs when they are within the communication range of an edge server. Since the physical distance between the edge server and the driver is very short, the data transmission time is commonly considered negligible. Consequently, the overall latency is approximated by the model's inference time for a single data point. 
	CC is roughly estimated by $\textit{Model Size} \times 2 \times \newline \textit{Number of communication round} \div \textit{Local epoch}$. \textit{Local epoch} = 1 for the comparative baselines. \textit{Local epoch} = 10 for MFed because the model's loss is the lowerest and the model converges the fast when \textit{Local epoch} = 10 as shown in Fig.~\ref{fig:FL_diff_localEpoch}.
	CL, MM, and PP refer to whether a framework is built with distributed collaborative learning, supports multimodal learning, and preserves data privacy (i.e., privacy leakage during parameter sharing).
	Besides, to more reliably evaluate the model's generalization performance on the test set and prevent overfitting, all our results were collected using K-fold cross-validation (with K=5).
	Our results are averaged across multiple runs and the variance is ±0.25.
}

\begin{table*}[!ht]
	\caption{MRHD evaluation results (\%) on our dataset.}
	\vspace{-4mm}
	\label{tab:MRHDEvaluationResults_own}
	\center
	\setlength\tabcolsep{1.5pt}
	\begin{tabular}{ccc}
		\toprule
		Model & Accuracy (\%) on our dataset   &  Accuracy (\%) on the CrisisMMD dataset   \\
		\midrule
		MobileNetV2~\cite{sandler2018mobilenetv2} & 83.33 & 85.32\\
		Bert~\cite{devlin2018bert} & 95.10 & 88.47\\
		\midrule
		Zou et al.~\cite{zou2021disaster}& 94.84 & 87.14  \\
		Wang et al.~\cite{wang2022hybrid}& 95.33 & 89.92 \\
		Choi et al.~\cite{choi2019embracenet}& 86.54 & 86.92  \\	
		Pranesh et al.~\cite{praneshexploring} & 85.90 & 85.81  \\
		Abavisani et al.~\cite{abavisani2020multimodal} & 98.08 & 91.78 \\
		\textbf{MRHD-noPretrain (ours)} &  97.79  & 86.93 \\
		\textbf{MRHD (ours)} & \textbf{99.14} & \textbf{92.00} \\
\bottomrule
\end{tabular}
\vspace{-1mm}
\end{table*}

\begin{figure}[!ht]
\centering
\includegraphics[width=0.4\textwidth]{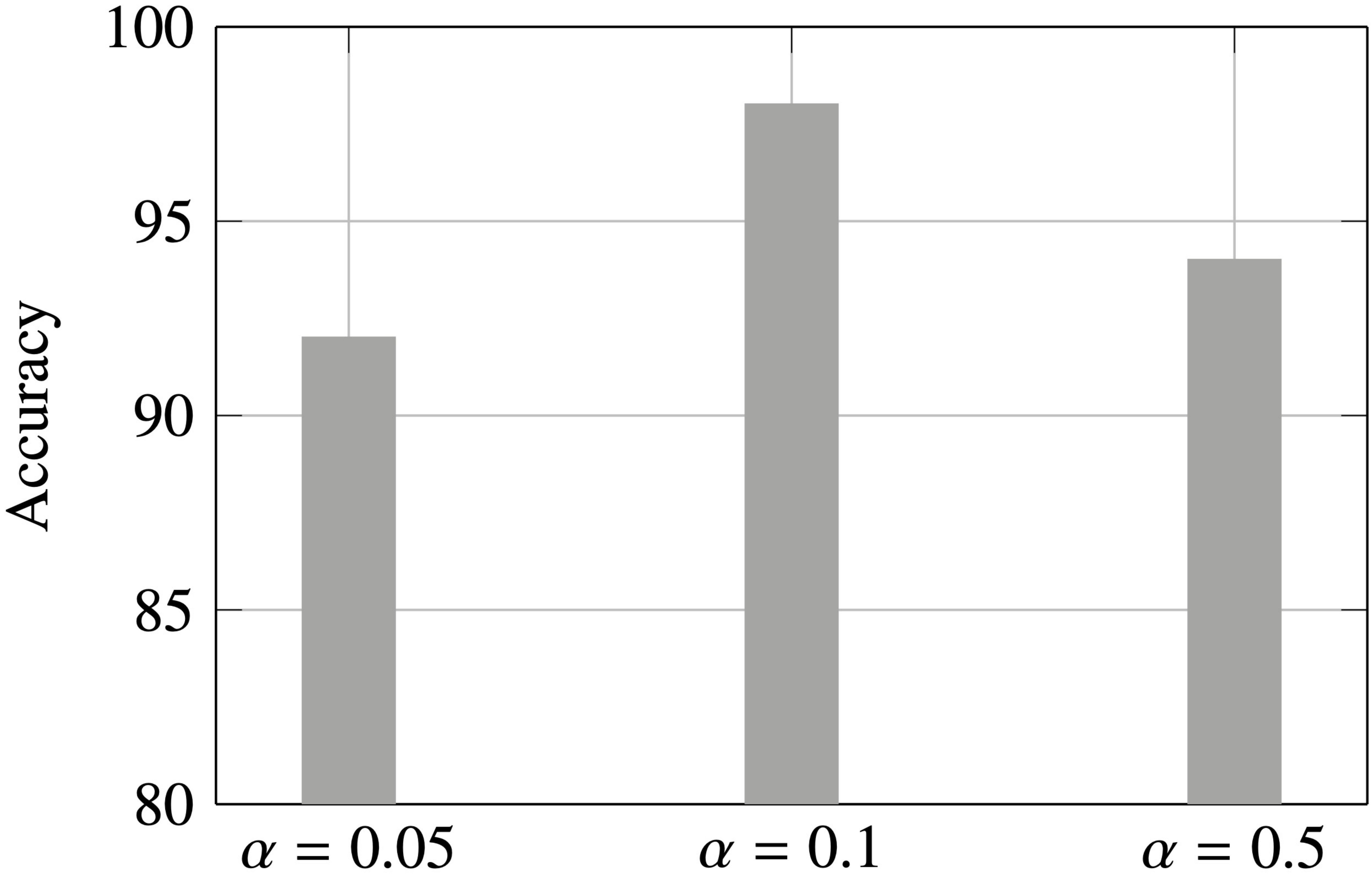}
\caption{\textcolor{black}{MRHD's accuracy over different $\alpha$.}}
\label{fig:different_alpha}
\end{figure}

\subsection{MRHD Results and Evaluation} \label{subsec:MRHDResultsandEvaluation}

\begin{figure*}[!ht]
\centering
\includegraphics[width=0.7\textwidth]{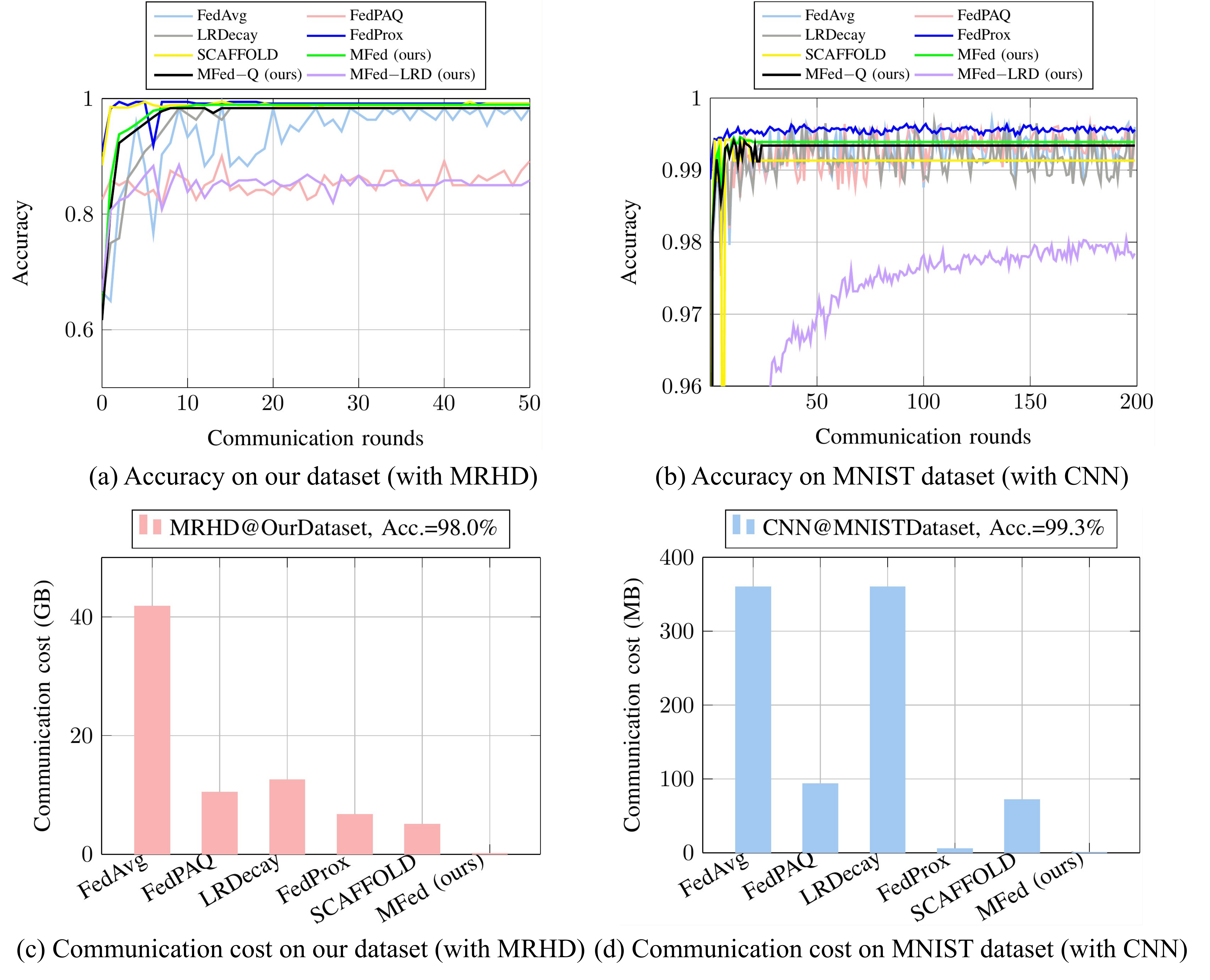}
\caption{\textcolor{black}{Performance comparison of MFed on both the road danger dataset and the MNIST dataset under non-i.i.d. setting I (4 classes per client)}.}
\label{fig:FL_performance}
\end{figure*}

\begin{figure*}[!ht]
\centering
\includegraphics[width=0.7\textwidth]{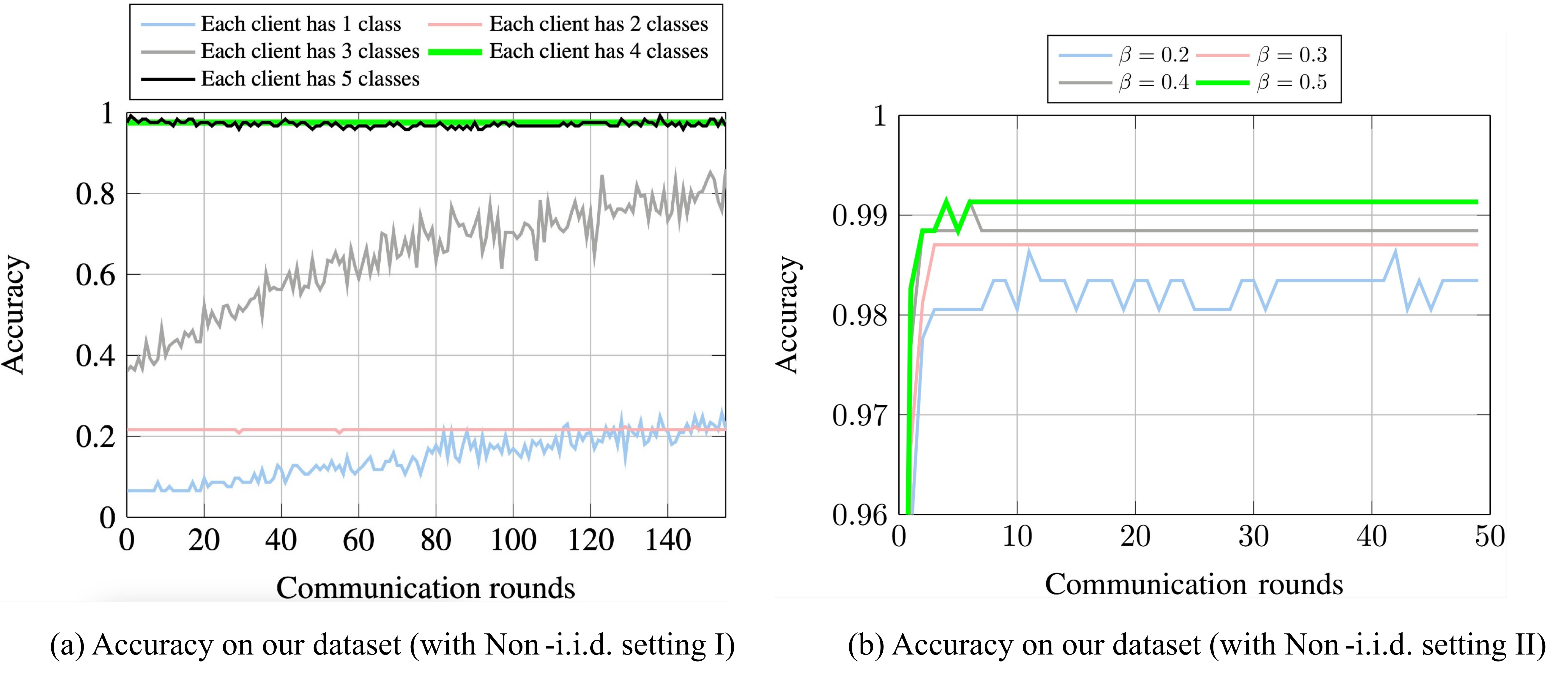}
\caption{\textcolor{black}{Performance comparison of MFed under different non-i.i.d. distributions on our dataset using the MRHD model.}}
\label{fig:distribution_own}
\end{figure*}

Table~\ref{tab:MRHDEvaluationResults_own} presents the classification accuracy of the proposed MRHD framework and several state-of-the-art baselines on two datasets: our custom dataset and the public CrisisMMD dataset. Across both settings, the full MRHD model consistently outper-forms all compared methods. On our dataset, MRHD achieves an accuracy of 99.14\%, outperforming the best baseline (Abavisani et al. [17], 98.08\%) by 1.06\%. Similarly, on the CrisisMMD dataset, MRHD reaches 92.00\% accuracy, surpassing the prior SOTA (Abavisani et al. \cite{abavisani2020multimodal}, 91.78\%) by a clear margin. To further validate the contribution of the pre-training strategy, we include an ablation variant, MRHD-noPretrain (when not employing the proposed triplet loss function). The results show that removing the pre-training component leads to a noticeable performance drop: from 99.14\% to 97.79\% on our dataset, and from 92.00\% to 86.93\% on CrisisMMD. This confirms that pre-training plays a critical role in boosting generalization, especially on the more challenging public dataset.

\textcolor{black}{As shown in Fig. \ref{fig:different_alpha}, we found that $\alpha=0.1$ yields optimal results compared to when $\alpha=0.05$ and $\alpha=0.5$.} This is reasonable because textual data includes high-level information, whereas visual data consists primarily of low-level information, and if the text-only part is not penalized, the overall loss would be excessively high.

\subsection{MFed Results and Evaluation}
\textcolor{black}{The findings illustrated in Fig.~\ref{fig:FL_performance} reveal that the accuracy of MFed surpasses FedAvg, LRDecay, and FedPAQ on our road danger dataset and the MNIST dataset. Although FedProx achieves similar high accuracy compared to MFed on our datasets, its communication cost (i.e., 6.68 GB) is higher than MFed on our datasets (i.e., 0.13 GB). Despite that FedProx achieves a slightly higher accuracy (i.e., 0.2\%) on MNIST dataset, it's communication cost (5.4 MB) is much higher than MFed (0.28 MB). SCAFFOLD and MFed have similar high performance, but SCAFFOLD's communication cost is much higher than MFed due to the fact that it uses a control variable with same dimension as the model to correct gradient biases. Hence, in its federated learning training, each client must transmit not only the trained model but also a control variable. The outstanding performance of MFed are stemmed from its integration of periodic parameter sharing, quantization, and learning rate decay strategies. Specifically, periodic parameter sharing largely reduces model transmission frequency (10 times lower), quantization significantly reduces model size (e.g., the model size of MRHD before and after quantization are 428 MB and 107 MB), and learning rate decay ensures fast convergence under non-i.i.d. heterogeneous real-world data (e.g., the model converges at 6/150 global communication round with/without applying learning rate decay strategy on MNIST dataset).
}

\textcolor{black}{Fig.~\ref{fig:FL_diff_localEpoch} demonstrates that as the number of global communication rounds increases, the loss function exhibits an overall declining trend despite occasional fluctuations across different local epochs. Here, "local epoch" refers to the number of training rounds performed by the client's local model on its local dataset before participating in federated training (i.e., before transmitting the local model to the federated parameter server). Generally, a larger local epoch reduces the required number of federated communication rounds, thereby lowering communication costs. However, an excessively large local epoch may adversely affect the overall performance of federated learning. Based on the above observations, we select a local epoch of 10, as at this value the global model converges relatively quickly while achieving a low loss (as shown in Fig.~\ref{fig:FL_diff_localEpoch}).
}

\textcolor{black}{We validate the performance of MFed under two different non-i.i.d. distributions (non-i.i.d. setting I and II in Table~\ref{tab:parameterSetup}) and compare the accuracy changes of MFed across different non-i.i.d. settings as the communication rounds increased (the results are shown in Fig.~\ref{fig:distribution_own}). The results in Fig.~\ref{fig:distribution_own} demonstrate that the higher the degree of non-i.i.d. in the client datasets, the lower the accuracy of MFed. For instance, as depicted in Fig.~\ref{fig:distribution_own}(a), when each client possesses samples from only 1 or 2 classes, MFed achieves an accuracy of merely around 20\%. However, as the number of classes per client increases (i.e., the degree of non-i.i.d. decreases), MFed reaches 80\% accuracy (when each client has 3 classes) and 99\% accuracy (when each client has 4 or 5 classes). Notably, when each client holds 4 or more classes (out of a total of 5 classes), the model accuracy shows no significant difference under federated averaging. Similarly, Fig.~\ref{fig:distribution_own}(b) shows that the test accuracy of MFed declines with decreasing $\beta$, which corresponds to an increase in data heterogeneity (non-i.i.d. rate). 
}

\begin{table*}[!ht]
\caption{\textcolor{black}{RoadFed evaluation results on our dataset}.}
\vspace{-4mm}
\label{tab:RoadFedEvaluationResults_ourDataset}
\center
\begin{tabular}{ccccccc}
\toprule
Framework & Accuracy  & Latency (\si{\second})  & CC (\si{GB}) & CL & MM & PP\\ 
\midrule
Edge-cloudRD~\cite{Edge-cloudRD} & 92.62  & \textbf{0.0015} & 0.29 &$\times$ &$\times$ & $\times$\\
EEFNet~\cite{EEFNet} & 89.90 & 0.0020  & 0.29 & $\times$ &$\times$ & $\times$ \\
Vondikakis et al.~\cite{vondikakis2024fedrsc}  & 85.42 &  0.028  &  1.71 & $\checkmark$ & $\times$ &$\times$ \\
Dwivedi et al.~\cite{dwivedi2024road} & 81.25  & 0.020  &  2.76 & $\checkmark$ & $\times$ &$\times$ \\
Saha et al.~\cite{saha2024federated} & 84.82  & 0.021  &  3.50  & $\checkmark$ & $\times$ &$\times$\\
Wu et al.~\cite{wu2024hierarchical} & 83.63 & 0.018   & 13.21  & $\checkmark$ &$\times$ & $\times$ \\
\midrule
RoadFed (ours) & \textbf{96.42}  & 0.0351  & \textbf{0.13} & $\checkmark$ & $\checkmark$ & $\checkmark$\\
\bottomrule
\end{tabular}
\vspace{-1mm}
\end{table*}

\begin{table*}[!ht]
\caption{\textcolor{black}{RoadFed evaluation results on CrisisMMD dataset}.}
\vspace{-4mm}
\label{tab:RoadFedEvaluationResults_CrisisMMDdataset}
\center
\begin{tabular}{ccccccc}
\toprule
Framework & Accuracy & Latency (\si{\second}) & CC (\si{GB}) & CL & MM & PP\\ 
\midrule
Edge-cloudRD~\cite{Edge-cloudRD} & 81.82   & \textbf{0.0012}   & 0.78 & $\times$ &$\times$ & $\times$\\
EEFNet~\cite{EEFNet} & 76.30 & 0.0022 & 0.39 & $\times$ &$\times$ & $\times$ \\
Vondikakis et al.~\cite{vondikakis2024fedrsc}  & 72.51  &  0.023  & 1.95  & $\checkmark$ & $\times$ &$\times$ \\
Dwivedi et al.~\cite{dwivedi2024road} & 71.84  &  0.024  & 2.22 & $\checkmark$ & $\times$ &$\times$ \\
Saha et al.~\cite{saha2024federated} & 72.95 &  0.022   & 2.26  & $\checkmark$ & $\times$ &$\times$\\
Wu et al.~\cite{wu2024hierarchical} & 67.63 & 0.024 & 13.10 & $\checkmark$ &$\times$ & $\times$ \\
\midrule
RoadFed (ours) & \textbf{90.28}  & 0.048  & \textbf{0.21} & $\checkmark$ & $\checkmark$ & $\checkmark$\\
\bottomrule
\end{tabular}
\vspace{-1mm}
\end{table*} 

\subsection{MLDP Results and Evaluation}
\begin{figure}[!ht]
\centering
\includegraphics[width=0.4\textwidth]{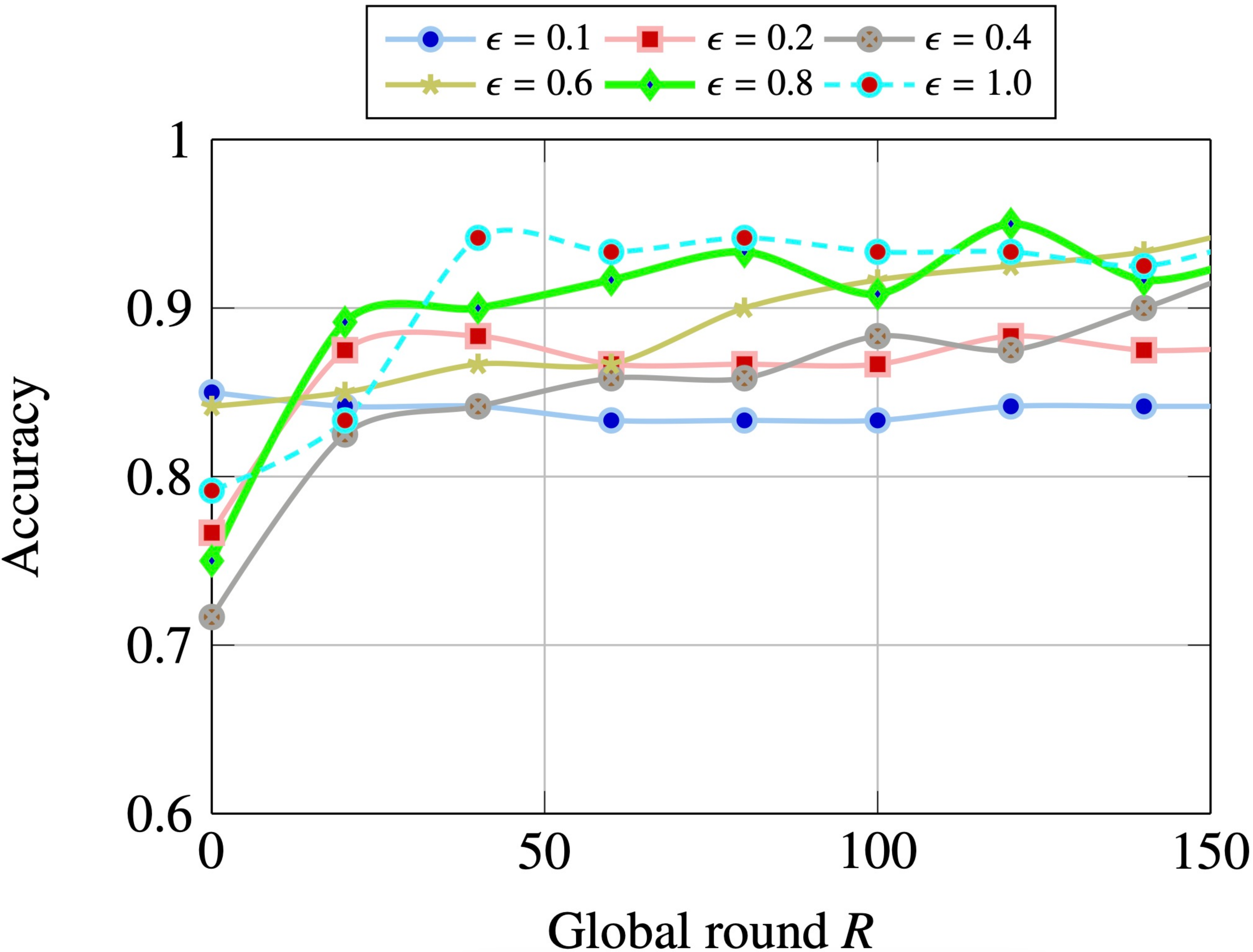}
\caption{The effect of $\epsilon$ in MLDP. $\epsilon = 0.8$ is selected as a trade-off between road hazard detection accuracy and data privacy preservation.} \label{fi:MLDPResults}
\end{figure}

The influence of the privacy budget $\epsilon$ of MLDP on the detection performance of RoadFed has been evaluated, with findings presented in Fig.~\ref{fi:MLDPResults}. \textcolor{black}{
According to \cite{dwork2014algorithmic}, differential privacy allocates $\epsilon$ privacy budget for a query. In our case, one query is equivalent to one data point (with dimension $d$) that is about to be transferred to an edge server. To ensure the total budget does not exceed $\epsilon$, a natural choice is to set $\epsilon_i=\epsilon/d$ for each dimension. For MLDP, the sensitivity of a data point often scales with its dimension $d$. By allocating $\epsilon/d$, the noise variance per dimension remains controlled, ensuring reasonable utility while satisfying the total $\epsilon$ constraint. Besides, if the entire $\epsilon$ were spent on a single dimension, other dimensions would lose protection, leading to poor utility in multivariate analysis. 
Therefore, in MLDP, for 1D text vectors (with the dimension of $d$), we allocate $\epsilon/d$ per dimension, while for 2D images (with the dimension of $dxd$), $\epsilon/d^2$ privacy budget per pixel is allocated to each data point. 
} 
\textcolor{black}{The required privacy budget is significantly reduced after using the proposed dimension reduction technique in MLDP, which avoids the problem of adding excessive differential privacy noise due to high data dimensionality, which would otherwise lead to a decrease in data utility.}

As depicted in Fig.~\ref{fi:MLDPResults}, the road danger detection model's accuracy remains below 90\% when $\epsilon$ is under 0.2, subsequently increasing rapidly as $\epsilon$ rises from 0.4 to 0.8. Specifically, compared to the detection accuracy at $\epsilon=0.001$, the accuracy of RoadFed with $\epsilon=0.8$ increases by 12.43\%. Additionally, the accuracy of the model when $\epsilon = 0.8$ is approximately 5\% and 1\% higher than that of when $\epsilon=0.4$ and $\epsilon=0.6$, respectively. There is minimal variation in RoadFed's detection accuracy when transitioning from $\epsilon = 0.8$ to $\epsilon = 0.1$. Hence, $\epsilon=0.8$ is chosen as the privacy level to strike a favorable balance between the detection performance of the local model and the privacy of the data.
\textcolor{black}{It is worth noting that the dimension reduction technique in MLDP also affects the utility-privacy trade-off. The reason behind it is that given an input image with dimensions of $d_1 \times d_2$, the privacy budget allocated to each dimension is $\epsilon/(d_1 \times d_2)$. To ensure privacy protection, $\epsilon$ must be reduced with the increase of the input image's dimensionality. A smaller $\epsilon$ means the addition of more noise, thereby reducing data utility. Therefore, when the image is downscaled to a dimension smaller than $1 \times 64$, the privacy budget $\epsilon$ that satisfies the utility-privacy trade-off will be smaller. Conversely, when the image is downscaled to a dimension larger than $1 \times 64$, the privacy budget $\epsilon$ that satisfies the utility-privacy trade-off will be larger.}

\begin{figure}[!ht]
\centering
\includegraphics[width=\linewidth]{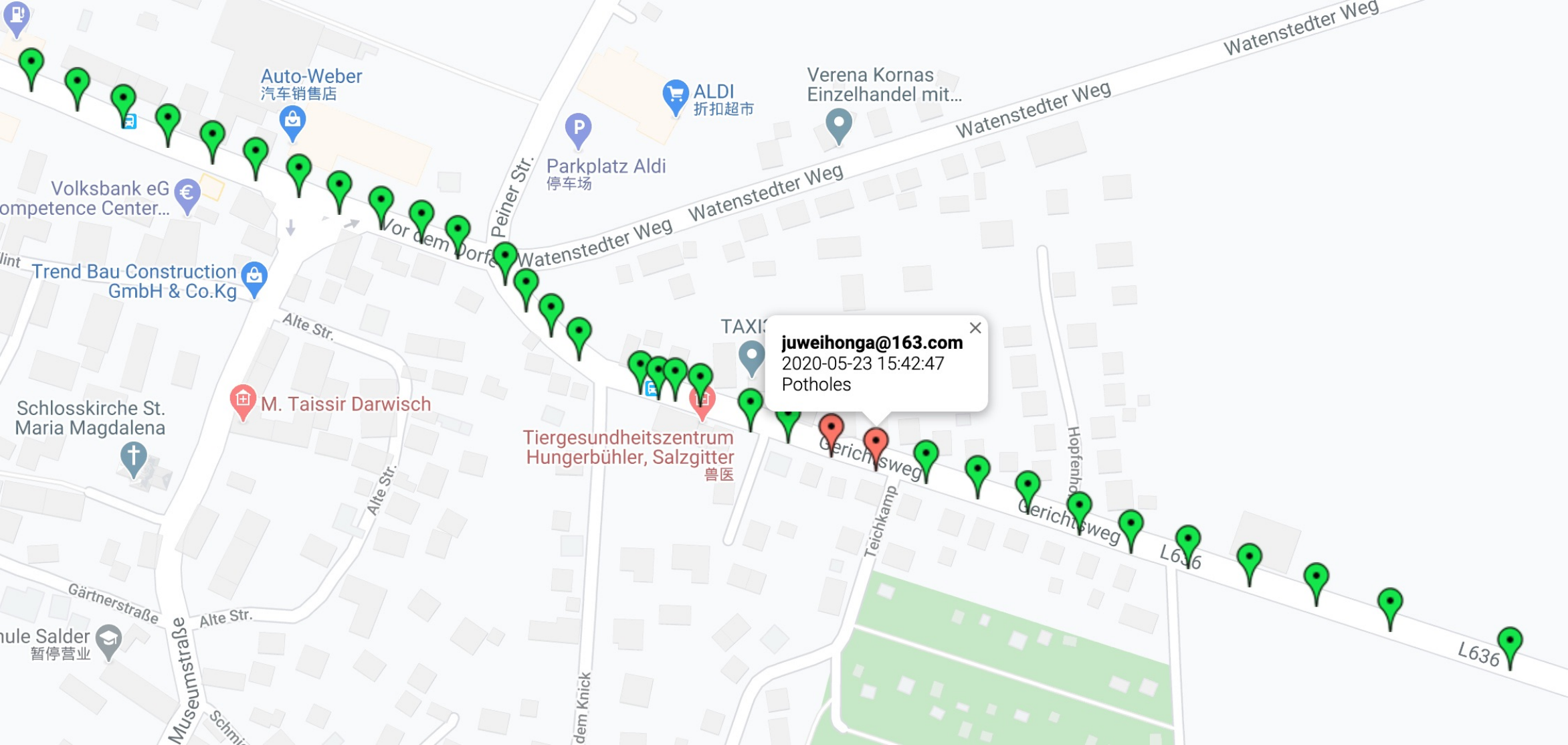}
\caption{Detection Result Display, where the red and green markers refer to road areas with and without road hazards. One can click the red markers to see details about the road hazards, including the type of road hazards and a timestamp.}
\label{fig:display}
\end{figure}

\subsection{RoadFed Framework Results and Evaluation}
\textcolor{black}{RoadFed is evaluated against Edge-cloudRD~\cite{Edge-cloudRD}, EdgeRD~\cite{saha2024federated}, EEFNet~\cite{EEFNet}, Vondikakis et al. \cite{vondikakis2024fedrsc}, Dwivedi et al. \cite{dwivedi2024road}, Saha et al. \cite{saha2024federated}, and Wu et al. \cite{wu2024hierarchical}. 
The performance metrics assessed include accuracy, F1-score, latency (s), Communication Cost (CC (GB)), Collaborative Learning (CL), Multimodal (MM), and Privacy-preserving (PP) for various frameworks, as detailed in Table~\ref{tab:RoadFedEvaluationResults_ourDataset} and Table~\ref{tab:RoadFedEvaluationResults_CrisisMMDdataset}. 
}

\textcolor{black}{On our dataset, as shown in Table~\ref{tab:RoadFedEvaluationResults_ourDataset}, RoadFed achieves the highest accuracy of 96.42\%, significantly outperforming the state-of-the-art, while also exhibiting competitive latency (0.0351 s). Notably, RoadFed achieves the lowest communication cost (CC = 0.13 GB), which is substantially lower than all other methods. Moreover, RoadFed uniquely supports collaborative learning (CL), multimodal learning (MM), and privacy preserving (PP), as indicated by the checkmarks in the table. Notably, Wu et al. \cite{wu2024hierarchical} has a remarkably high communication cost due to it's two-layer hierarchical structure of FL, which means the model parameters that need to be transmitted are twice that of a typical FL setup, even if the convergence speed remains the same.}

\textcolor{black}{On the CrisisMMD dataset, as illustrated in Table~\ref{tab:RoadFedEvaluationResults_CrisisMMDdataset}, RoadFed again achieves the highest accuracy (90.28\%), surpassing the comparative frameworks. RoadFed’s communication cost (0.21 GB) is the lowest compared with the baselines. Although its latency (0.048 s) is higher than Edge-cloudRD (0.0012 s), it remains within acceptable bounds for real-time deployment. Importantly, RoadFed is the only framework that supports CL, MM, and PP on this dataset as well, highlighting its versatility and adaptability to diverse deployment scenarios.}

\textcolor{black}{Overall, RoadFed has around 4\% and 9\% improvement in accuracy compared to the state-of-the-art on our self-collected dataset and the CrisisMMD dataset. Additionally, RoadFed has 0.035 s and 0.0082 s latency on our dataset and the CrisisMMD dataset. 
RoadFed has the lowest communication cost compared with the existing collaborative learning-based frameworks, and the underlying reason behind this is that it converges faster on non-i.i.d. data stemming from the use of learning rate decay strategy, its model size is smaller due to quantization, and the model transmission frequency is lower because of the periodic parameter sharing scheme. Overall, compared to the existing distributed systems, RoadFed offers a higher detection accuracy on both datasets, supports multi-modal data, and protects data privacy. 
}

\textcolor{black}{Additionally, the dataset we used contains a wide range of samples from various scenarios, including both urban and rural settings. It also features diverse weather conditions such as cloudy, rainy, snowy, and sunny days, different levels of illumination (very dark/light), and various types of occlusions and obstructions (e.g., vehicles and pedestrians). The data distribution is also highly imbalanced. Achieving a high accuracy of 96.42\% on such challenging datasets demonstrates the remarkable robustness of RoadFed.
Additionally, the experimental results in Fig.~\ref{fig:FL_performance} (a), Fig.~\ref{fig:FL_performance} (b), and Fig.~\ref{fig:FL_diff_localEpoch} demonstrate that the proposed framework satisfies the design goal of fast convergence. Specifically, the experimental results in Fig.~\ref{fig:FL_performance} (a), Fig.~\ref{fig:FL_performance} (b) demonstrate that the proposed RoadFed converges much faster than existing federated learning frameworks. The results in Fig.~\ref{fig:FL_diff_localEpoch} show that the proposed MFed converges fast under different local epochs. 
}

\subsection{Displaying the road danger detection results on Google Maps}
After identifying potential hazards on the road, alerts regarding these risks (including their danger types, GPS coordinates, and a timestamp) are forwarded to the edge server. The server displays them on Google Maps. 
It is dynamically refreshed whenever new road hazards are encountered.
Armed with this hazard map, road users can capture the current condition of the road network and determine the most secure routes for their journeys while road management authorities can provide timely road maintenance.
A dedicated webpage has been crafted to present the detected road hazards. As depicted in Fig. \ref{fig:display}, the detected road hazards are seamlessly integrated onto Google Maps. 
In the map, green GPS markers signify areas that are deemed safe (indicating no detected hazards), whereas red GPS markers highlight sections that are considered hazardous (suggesting the presence of one or multiple hazards). 
By clicking on a red marker, one can see details about the identified dangers, including the type of the detected road danger and a timestamp.

\section{Discussion} \label{sec:Discussion}
\subsection{\textcolor{black}{Handling Missing Modalities}}
\begin{figure}[!ht]
\centering
\includegraphics[width=0.45\textwidth]{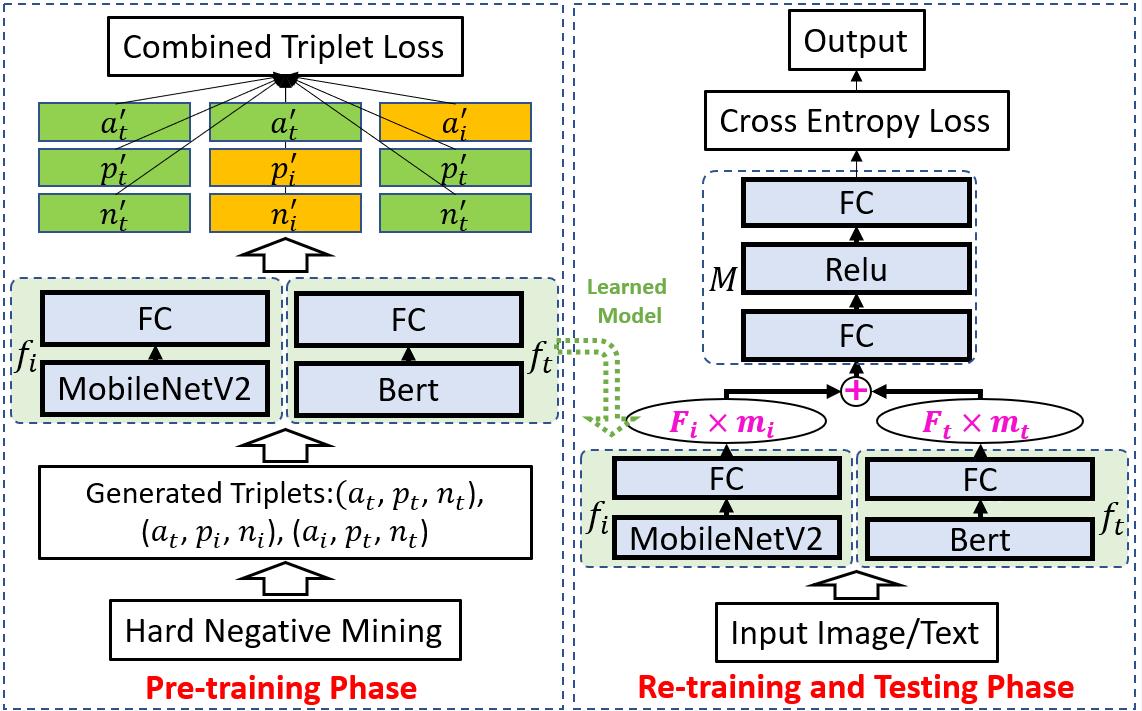}
\caption{\textcolor{black}{The introduced Multimodal Road Hazard Detector that uses a masking mechanism (marked in magenta color) to handle missing modalities.}} \label{fi:MissingModality}
\end{figure}

\textcolor{black}{To handle missing modalities (e.g., only image or only text is available during the training and inferencing phases), practitioners can simply apply a masking mechanism after the feature extractors (MobileNetV2/BERT) before aggregation. As depicted in Fig.~\ref{fi:MissingModality}, each modality is associated with a binary mask $m$ (0 for missing, 1 for present), i.e., $m_i$ for image modality and $m_t$ for text modality. The mask $m$ is multiplied with its corresponding feature vectors ($F_i$ for image modality and $F_t$ for text modality). The image branch output becomes $F_i \times m_i$, and the text branch output becomes $F_t \times m_t$. The input of the FC layers thus becomes $F_i \times m_i + F_t \times m_t$. During training, practitioners can randomly drop one or both modalities with a certain probability (e.g., 5\% per modality) to simulate missing data. The mask is applied to zero-out the features of the dropped modality, teaching the model to rely on available modalities and improving robustness. During online inference, if a modality is missing (e.g., no text input), its mask is set to 0. The model dynamically adjusts based on the masks. This approach is effective because it operates on the feature level rather than the raw input, preserving the pre-trained representations.}

\textcolor{black}{Table~\ref{tab:missingModalities} shows the performance of RoadFed when encountering missing modalities. The results demonstrate that with a simple masking mechanism, RoadFed performs well when some road users only upload single modality data (i.e., >90\% accuracy on our dataset and >85\% accuracy on the CrisisMMD dataset).}

\begin{table}[!ht]
	\caption{\textcolor{black}{RoadFed evaluation results in case of missing modalities}.}
	\vspace{-4mm}
	\label{tab:missingModalities}
	\center
	\setlength\tabcolsep{1.5pt}
	\begin{tabular}{ccc}
		\toprule
		Datasets & Missed Modalities & Accuracy \\ 
		\midrule 
		Self-collected & 15\% image & 93.27 \\
		& 15\% text & 90.84  \\
		& 15\% image and 15\% text & 90.35 \\
		& No missing modalities & 96.42 \\
		\midrule
		CrisisMMD & 15\% image & 89.02 \\
		& 15\% text & 86.52 \\
		& 15\% image and 15\% text & 85.29 \\
		& No missing modalities & 91.51 \\
		\bottomrule
	\end{tabular}
	\vspace{-1mm}
\end{table}


\subsection{\textcolor{black}{Latency on Typical Edge Servers}}
\textcolor{black}{As the training and testing of the MRHD model in RoadFed are performed on edge servers, except for Laptop (with a RTX 3090 GPU), we also evaluate its inference latency on other edge devices (e.g., Jetson AGX Orin and Jetson Orin NX) to verify its practicality under typical edge servers. The results in Table~\ref{tab:latency} show that on a high-performance edge server (e.g., Jetson AGX Orin 64G), latency reaches approximately 0.194 s, while on a resource-constrained embedded edge device (e.g., Jetson Orin NX), it is about 0.375 s. These measurements confirm that the model can also achieve reasonably low inference latency on common edge servers with GPUs, demonstrating its feasibility for real-world edge deployment.}

\begin{table}[!ht]
\caption{\textcolor{black}{The latency of RoadFed on different edge servers}.}
\vspace{-4mm}
\label{tab:latency}
\center
\setlength\tabcolsep{1.5pt}
\begin{tabular}{cc}
\toprule
Edge Hardware & Latency (s) \\ 
\midrule 
Jetson AGX Orin 64GB & 0.194\\
Jetson Orin NX 16GB & 0.375 \\
\bottomrule
\end{tabular}
\vspace{-1mm}
\end{table} 

\subsection{\textcolor{black}{Participant Scalability Analysis}}
\begin{figure}[t]
\centering
\includegraphics[width=0.4\textwidth]{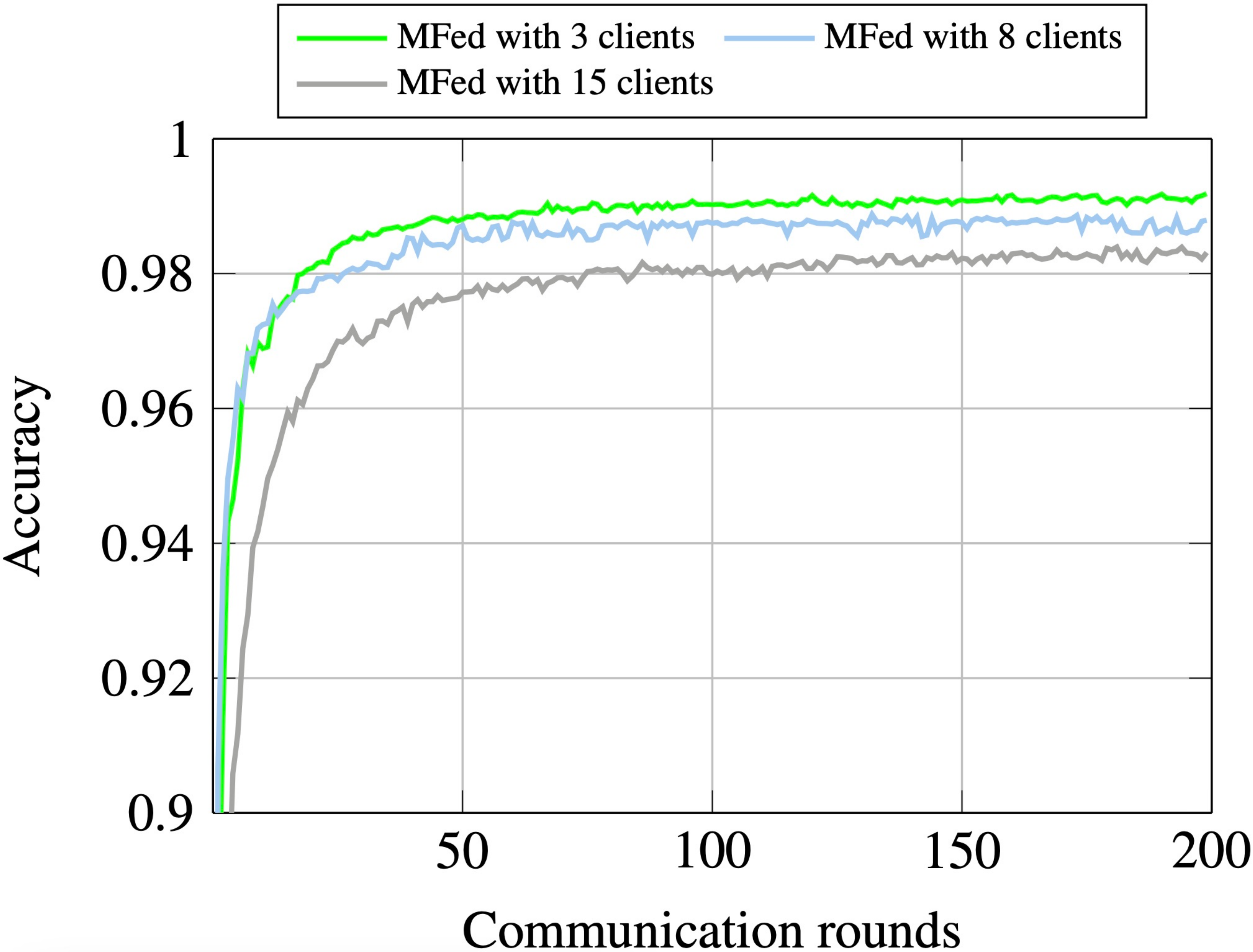}
\caption{\textcolor{black}{Performance comparison of RoadFed with different numbers of clients on the MNIST dataset using the CNN model.}}
\label{fig:mnist_with_different_clients}
\end{figure}

\textcolor{black}{To investigate the impact of varying FL client numbers on RoadFed's performance, we compare the model accuracy for 3, 8, and 15 clients, with the results presented in Fig.~\ref{fig:mnist_with_different_clients}. As shown in the figure, the performance of MFed generally decreases as the number of clients increases. The reason behind this observation is that under the constraint of a fixed total sample size, an increase in the number of clients leads to a reduction in the average number of samples per client. This local data scarcity can lower the training quality and stability of local models, resulting in less accurate gradient updates being sent to the server and ultimately affecting the global model's accuracy. Besides, as the number of clients grows, the data distribution for each individual client becomes more skewed. This severe data skew intensifies the non-i.i.d. problem, making it difficult for the global model to effectively aggregate local updates from all clients, which in turn impacts model performance.}

\section{Conclusion} \label{sec:conclusion}
We introduce the RoadFed framework for cost-effective, efficient, and private detection and alerting of road hazards. In RoadFed, we present a Multimodal Road Hazard Detector that incorporates a new loss function that considers inter-class and intra-class correlation to enhance the classification of road hazards from different data modalities (i.e., images and texts). An effective FL method is also designed to improve the accuracy of local road hazard detection models on the edge servers, drastically minimizing communication and computational expenses while ensuring model convergence. A multimodal LDP-based scheme is proposed to safeguard private information before transmitting it to the edge servers. This method effectively addresses the high dimensionality challenges associated with LDP. Experimental outcomes show that RoadFed can rapidly respond to road hazards, achieving high accuracy with minimal communication costs while protecting data privacy. The proposed framework is well-suited for integration into advanced cooperative ITSs. Specifically, RoadFed can alert drivers and pedestrians of impending hazards, providing details and locations to help prevent accidents. It can offer dynamic route guidance to improve travel times, contributing to environmental benefits. Alerts about collisions, breakdowns ahead, and adverse road conditions because of weather, such as icy roads, can also be provided. Ultimately, road administration authorities can focus on areas with statistically higher occurrences of collisions and incidents. The proposed framework enhances ITSs’ data collection, storage, and analysis capabilities, supporting future policy development and improving traffic management. As part of the future work, we also plan to explore a more decentralized or redundant cloud architecture, which could enhance system reliability while still maintaining the same level of privacy protection for the clients.

\section*{Acknowledgment}
The authors would like to thank MD Samiur Rahman for their help in the experiments, and Weihong Ju for his assistance in building the web platform based on Google Maps for displaying the detection results.
    This work was supported in part by the National Natural Science Foundation of China (62406215, 72402156, 62072321), the fundamental Research Funds for the Central Universities, the Science and Technology Program of Jiangsu Province (BZ2024062), the Natural Science Foundation of the Jiangsu Higher Education Institutions of China (22KJA520007), Suzhou Planning Project of Science and Technology (2023ss03, SYG2025129, SNG2025010), and Key Laboratory of General Artificial Intelligence and Large Models in Provincial Universities, Soochow University.

\bibliographystyle{elsarticle-num} 
\bibliography{main}
\end{sloppypar}
\end{document}